\documentclass[useAMS,usenatbib]{mn2e}
\usepackage{hyperref}
\usepackage{graphicx}
 
\begin{document}

\title[ The star-formation rate density from z = 1-6]{
The star-formation rate density from z = 1-6
\thanks{{\it Herschel} is an ESA space observatory with science
instruments provided by European-led Principal Investigator
consortia and with important participation from NASA.}
}
\author[Rowan-Robinson M. et al]{Michael Rowan-Robinson$^{1}$
, Seb Oliver$^{2}$, Lingyu Wang$^{3}$, Duncan Farrah$^{4}$, 
\newauthor
David L. Clements$^{1}$, Carlotta Gruppioni$^{5}$, Lucia Marchetti$^{6}$, Dimitra Rigopoulou$^{7}$,
\newauthor
Mattia Vaccari$^{8}$\\
$^{1}$Astrophysics Group, Imperial College London, Blackett Laboratory, Prince Consort Road, London SW7 2AZ, UK\\
$^{2}$Astronomy Centre, Dept. of Physics \& Astronomy, University of Sussex, Brighton BN1 9QH, UK,\\
$^{3}$Department of Physics, Durham University, South Rd, Durham DH1 3LE, UK,\\
$^{4}$Department of Physics, Virginia Tech, Blacksburg, VA 24061, USA\\
$^{5}$INAF,  - Osservatorio Astronomico di Bologna, via Ranzani 1, I-40127 Bologna, Italy\\
$^{6}$Department of Physical Science, The Open University, Milton Keynes MK7 6AA, UK\\
$^{7}$Department of Astrophysics, University of Oxford, Keble Rd, Oxford OX1 3RH\\ 
$^{8}$Astrophysics Group, University of the Western Cape, Private Bag X17, 7535, Bellville, Cape Town, South Africa\\
}
\maketitle
\begin{abstract}
We use 3035 Herschel-SPIRE 500$\mu$m sources from 20.3 sq deg of sky in the HerMES Lockman, ES1 and XMM-LSS areas 
to estimate the star-formation rate density at z = 1-6.  
500 $\mu$m sources are associated first with 350 and 250 $\mu$m sources, and then with Spitzer 24 $\mu$m sources
from the SWIRE photometric redshift catalogue.  The infrared and submillimetre data are fitted with a set of 
radiative-transfer templates corresponding to cirrus (quiescent) and starburst galaxies.  Lensing candidates are
removed via a set of colour-colour and colour-redshift constraints.  Star-formation rates are found to extend from
$<$ 1 to 20,000 $M_{\odot} yr^{-1}$.  Such high values were also seen in the all-sky IRAS Faint Source Survey.
Star-formation rate functions are derived in a series of redshift bins from 0-6, combined with earlier far-infrared
estimates, where available, and fitted with a Saunders et al (1990) functional form.  The star-formation-rate density as a function
of redshift is derived and compared with other estimates.  There is reasonable agreement with both infrared and ultraviolet
estimates for z $<$ 3, but we find higher star-formation-rate densities than ultraviolet estimates
at z = 3-6.  Given the considerable uncertainties in the submillimetre estimates, we can not rule out the possibility
that the ultraviolet estimates are correct.  But the possibility that the ultraviolet estimates have seriously underestimated
the contribution of dust-shrouded star-formation can also not be excluded.

\end{abstract}


\section{Introduction}

The history of the determination of the evolution of the integrated star-formation rate density
has been controversial.  Lilley at al (1996) and Madau et al (1996) gave estimates based purely on the
ultraviolet (uv) light from galaxies, with a correction for extinction based on a screen model.
Rowan-Robinson et al (1997) showed from ISO data that these estimates were likely to be significantly 
underestimated. More recent uv surveys (e.g. Wyder et al 2005,
Schiminovich et al 2005, Dahlen et al 2007, Reddy and Steidel (2009), Cucciati et al 2012) and 
infrared (ir) surveys (eg Sanders et al 2003, Takeuchi et al 2003,
Magnelli et al 2011, 2013, Gruppioni et al 2013) are now in reasonable agreement for z $<$ 3.
The Gruppioni et al (2013) study of Herschel sources at 70, 100 and 160 $\mu$m is especially
significant in capturing the total far infrared luminosity, and hence a more accurate estimate of the star-
formation rate.  
Madau and Dickinson (2014) have given a comprehensive review of the current
situation. 

At higher redshifts (z $\sim$ 4-10) we have only ultraviolet estimates (Bouwens et al 2012a,b, 
Schenker et al 2013) and so the problem remains: is the contribution of dust-shrouded
star-formation being properly accounted for?  The problem can be seen clearly by imagining
external observations of our own Galaxy.  The blue and ultraviolet light would be dominated
by young stars which would be subject to an average extinction of a few tenths of a magnitude,
due to the dust spread through the interstellar medium.  The infrared emission from this
optically thin dust makes up the infrared ‘cirrus’.  However the contribution of newly formed
stars embedded in dense molecular clouds would not be accounted for.  In the case of our
Galaxy this would result in an underestimate of the total star-formation rate by only about
10$\%$, but for luminous starbursts the underestimate could be over a factor of 100. 

To address this question we really need to analyse the star-formation history separately
for relatively quiescent galaxies like our own and for starburst galaxies.  In the latter
star formation appears to be driven primarily by major mergers whereas for quiescent
galaxies the driver is either interaction with companions or infall of gas from the cosmic web
(Daddi et al 2010, Rodighiero et al 2011).

The {\it Herschel} mission (Pilbratt et al 2010) gives us for the first time samples of galaxies 
for which we have the
full ultraviolet to submillimetre spectral energy distributions (SEDs) reaching out to z = 6
and so allows us to explore the evolution of the star-formation rate density beyond z = 3.
The samples we focus on in this paper are from the {\it HerMES} survey (Oliver et al 2012),
made with the SPIRE sub-millimetre camera (Griffin et al 2010) and
are selected at 500 $\mu$m, where the negative K-correction gives us greater visibility of 
the high redshift universe than selection at 250 or 350 $\mu$m (Franceschini et al 1991).  
500 $\mu$m selection also has the benefit
that in the majority of cases we have 250 and 350 $\mu$m data, which give us valuable
spectral energy distribution (SED) information.
Rowan-Robinson et al (2014) have given a very detailed discussion of the 
SEDs of a sample of 957 galaxies in the HerMES-Lockman area.  Here we extend the sample to the ES1 and XMM-LSS
areas of the HerMES survey.

The structure of this paper is as follows: in section 2 we define our samples and describe the 
determination of the star-formation rate via modelling of the spectral energy distributions 
(SEDs). In section 3 we derive the star-formation rate function in redshift bins from z = 0-6.  
In section 4 we use this information to determine the evolution of the star-formation rate 
density.  Section 5 gives our discussion and conclusions.

A cosmological model with $\Lambda$=0.7, $h_0$=0.72 has been used throughout.

\begin{table*}
\caption{Numbers of 500 $\mu$m sources in HerMES fields}
\begin{tabular}{llllll}
field & area in common & no. of 500 $\mu$m & candidate & no. of 500 $\mu$m sources & limiting 500 $\mu$m\\
& with SWIRE (sq deg) & sources & lenses & not associated with SWIRE & flux-density (mJy)\\
&&&&&\\
Lockman  & 7.5 & 957 & 109 & 368 & 25 \\
&&&&&\\
ES1 & 3.8 & 478 & 48 & 205 & 25\\
&&&&\\
XMM-LSS & 9 & 767 & 72 & 259 & 30 \\
&&&&&\\
EN1 & 3.1 & 116 & 23 & 5 & 40\\
&&&&&\\
CDFS & 2.9 & 129 & 17 & 22 & 40\\
\end{tabular}
\end{table*}

\section{Determination of star-formation rate through SED modelling}

The Lockman sample has been studied in detail by Rowan-Robinson et al (2014).  Our starting point
was the HerMES SPIRE (SCAT) 500 $\mu$m catalogue (Wang et al 2014a) with sources detected at 500 $\mu$m
without using any prior information from other SWIRE bands (i.e. ‘blind’ 500 $\mu$m catalogue).
500 $\mu$m sources from the HerMES survey in Lockman were associated first with 350 $\mu$m sources,
and only accepted as credible if there was a good (5-$\sigma$) 350 $\mu$m detection, and then
with 250 $\mu$m sources
and with galaxies from the SWIRE photometric redshift catalogue (Rowan-Robinson et al
2013), using automated SED fits to the SWIRE data to select the most likely 24 $\mu$m association.
There are then 957 sources for which we have SEDs from 0.36 to 500 $\mu$m, and a
further 368 500-350-250 $\mu$m sources with no SWIRE counterpart.  The 957 sources with
SWIRE counterparts were modelled with an automatic infrared template-fitting routine
using a set of standard templates: M82 and A220 starbursts, a young starburst template,
normal and cool cirrus templates, and an AGN dust torus template.  These templates have
been derived from full radiative-transfer treatments (Efstathiou et al 2000, Efstathiou 
and Rowan-Robinson 2003, Rowan-Robinson 1995).  
Full details of the templates used are given via a readme page
\footnote{\url{http://astro.ic.ac.uk/public/mrr/swirephotzcat/templates/readme}}.

109 lensing candidates were identified through their anomalous SEDs, which in Rowan-Robinson et al (2013)
we characterised by a set of colour-colour
and colour-redshift constraints.  These lensing candidates have been removed from the present study, leaving a
sample 0f 848 submillimetre galaxies with SWIRE counterparts.

We have applied a similar analysis to the ES1, EN1, XMM-LSS and CDFS areas within the HerMES
survey.  A detailed discussion of the HerMES catalogues, and their completeness and reliability
is given by Wang et al (2014a).  Full details of the SWIRE photometric redshift catalogue in these areas are given via
a readme page \footnote{\url{http://astro.ic.ac.uk/public/mrr/swirephotzcat/readmeSWIRErev}}, including details of 
reprocessing of the XMM-LSS and ES1 areas to take advantage of new photometry.
Table 1 summarises the areas and numbers of sources for each region.  Plots of S24 versus 
S500 for each region (not shown here) allow an estimate of the 500 $\mu$m completeness limit for
each region. Because we require sources to be detected at both 500 and 350 $\mu$m at 5 times the total
noise (confusion plus instrumental), our completeness limits are higher than the confusion limits
quoted by Nguyen et al (2010), Wang et al (2014a).
Only ES1 and XMM-LSS are of comparable depth to Lockman ($\sim$25 mJy): these regions have been added to
the present study, yielding a total area covered of 20.3 sq deg. The EN1 and CDFS areas do not
play a part in the remainder of the paper.

Because our detection threshold at 500 $\mu$m is set at 5-$\sigma$, where $\sigma$ is the total (confusion
plus instrumental) noise, we do not believe our
sample is seriously compromised by completeness or confusion issues.  An extensive discussion
of confusion and misassociation issues was given by Rowan-Robinson et al (2014). The chance of
misassociation with the wrong 24 $\mu$m counterpart was estimated as 5$\%$.

\begin{figure*}
\includegraphics[width=7cm]{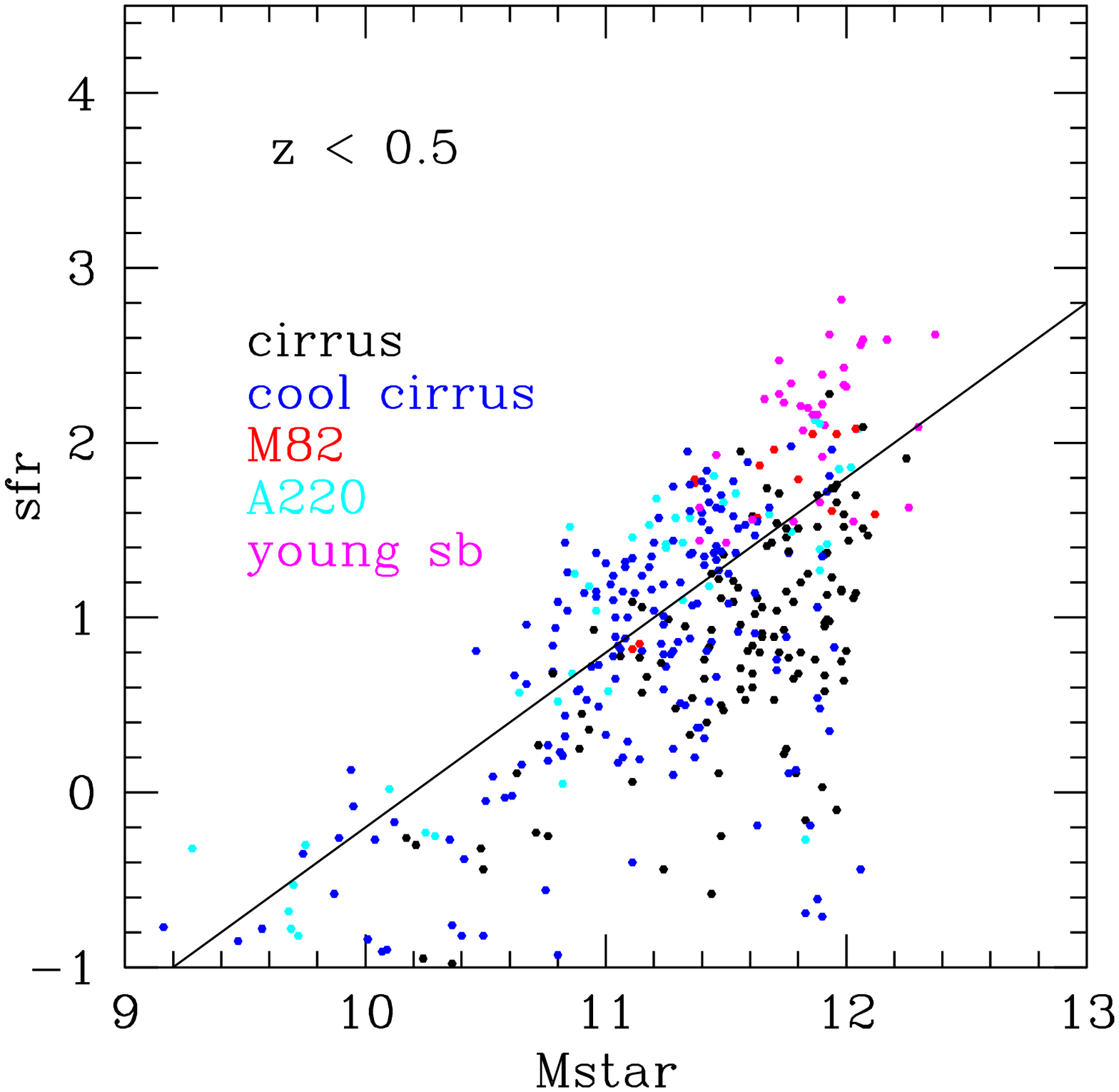}
\includegraphics[width=7cm]{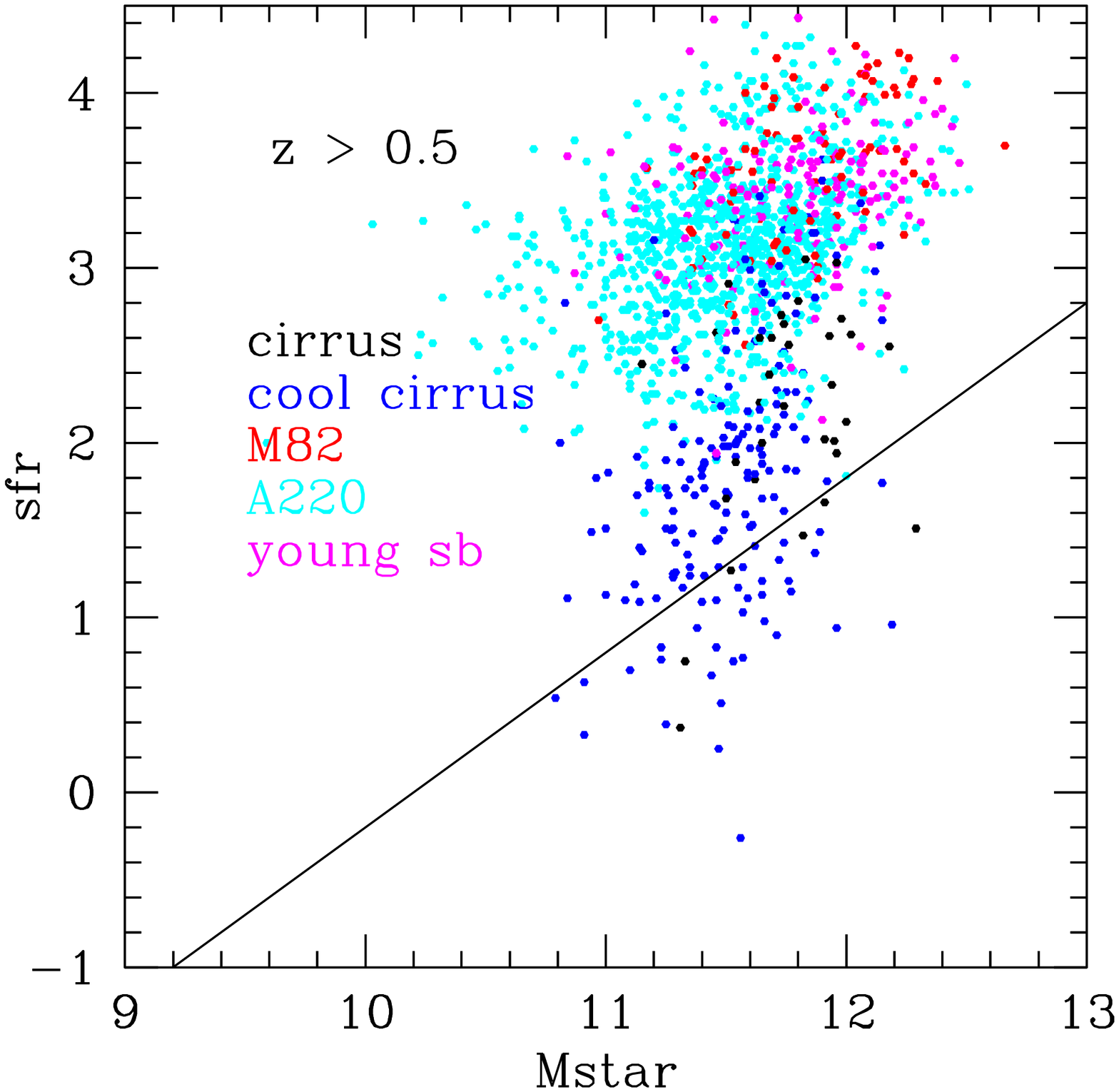}
\caption{
Star formation rate versus stellar mass for 1983 unlensed Lockman+XMM+ES1-SWIRE sources, colour-coded by
the dominant SED template component, for z$<$0.5 (L) and z$>$0.5 (R).  
There is a reasonably clear distinction between quiescent
galaxies, modelled with cirrus templates and with star-formation rate $<$ 100 $M_{\odot} yr^{-1}$,
and starburst galaxies, with star-formation rates greater than this.
An indicative line of slope 1 is plotted through the z$<$0.5 points.
}
\end{figure*}

As the templates used are derived from full radiative transfer models they allow the
star-formation rate to be estimated for each galaxy. The conversion factors from the infrared
luminosity have been given by Rowan-Robinson et al (2008). The models are spherically symmetric and
do not account, for example, for the possibility of multiple starburst locations in a galaxy.
In the case of cirrus components,
where optically thin dust is heated by starlight, we have made a correction here for the contribution
of old stars to the heating.  To do this we use the approximation noted by Rowan-Robinson (2003),
that the optical-nir SEDs of Hubble sequence galaxies can be modelled as a superposition of an 
elliptical galaxy template, representing stars older than 1 Gyr, and a residual component due to
young stars which varies with Hubble type.  We use the flux at 1.25 $\mu$m as a measure of the
old star component. 

The fraction of infrared radiation contributed by young stars in a galaxy of type t, y(t), is then estimated as
\begin{equation}
y(t)=1-\frac{QL(E))}{QL(t)}
%
%
\end{equation}

\noindent where
\begin{equation}
QL(t)=\int\frac{Q_{\nu,abs}f_{\nu} d\nu}{\nu f_{\nu}|_{1.25\mu m}}
\end{equation}

and $Q_{\nu,abs}$ is the absorption efficiency of the dust grains.

The star formation rate for cirrus components is then calculated from $y L_{cirr}$.
Table 2 gives the calculated values of y as a function of the optical template type.

\begin{table*}
\caption{Fraction of infrared (1-1000$\mu$m) emission contributed by young stars}
\begin{tabular}{llllllllllll}
&&&&&&&&&&&\\
j2  & 1 & 2 & 3 & 4 & 5 & 6 & 7 & 8 & 9 & 10 & 11 \\
&&&&&&&&&&&\\
type & E & E’ & Sab & & Sbc & & Scd & & Sdm & & sb \\
&&&&&&&&&&&\\
y & 0.0 & 0.0 & 0.360 & 0.288 & 0.216 & 0.412 & 0.607 & 0.752 & 0.896 & 0.866 & 0.835\\
\end{tabular}
\end{table*}

The optical templates used for the
SWIRE photometric redshift catalogue were fitted with stellar synthesis codes and so give estimates
of the stellar mass (Rowan-Robinson et al 2008).  Figure 1 shows the
star formation rate versus stellar mass for 1983 unlensed Lockman+XMM+ES1-SWIRE sources, colour-coded by
the dominant SED template component, divided into z $<$ 0.5 (L) and z $>$ 0.5 (R).  There is a reasonably 
clear distinction in Fig 1 between quiescent
galaxies, modelled with cirrus templates and with star-formation rate $<$ 100 $M_{\odot} yr^{-1}$,
and starburst galaxies, with star-formation rates greater than this.  However our distinction
between starburst and quiescent galaxies is based on SED type not on location in the sfr-$M_*$
diagram, unlike the $\lq$main-sequence$\rq$ locus (Brinchmann et al 2004, Daddi et al 2007, 2010,
Elbaz et al 2007, Noeske et al 2007, Genzel et al 2010, Rodighiero et al 2014).
32 of the 848 unlensed Lockman galaxies are fitted with a QSO template at 0.36-4.5 $\mu$m and do not feature
in Fig 1.

Figure 2L shows the star-formation rate versus redshift for Lockman+XMM+ES1 sources, with loci showing 
the selection imposed by the 500 $\mu$m sensitivity limit of 25 mJy for different template types. Galaxies of 
a certain type can fall below the limit imposed by the 500 $\mu$m limit for that type because SEDs can be
modelled with a mixture of template types, which contribute to different wavelength ranges.  We see 
that the selection loci for M82 starbursts and for
young starbursts are almost identical.  Of the 848 unlensed Lockman galaxies, 255 use an AGN dust torus template
in the far infrared and submillimetre SED fit ( and these do not contribute to the star-formation rate) but
for only two objects is an AGN dust torus the dominant contribution to the infrared luminosity.

We have also estimated photometric redshifts for the unidentified 500 $\mu$m sources from the 250-500
$\mu$m fluxes, allowing M82, Arp220 or young starburst templates, and these are shown as small filled 
circles in Fig 2L.  Redshifts for these generally range from z = 2-6. The redshift distribution for both
(unlensed) identified and unidentified sources in Lockman+XMM+ES1 is shown in Fig 3.  60$\%$ of our redshifts from 2-4
and almost all those at z$>$4 are unidentified sources.  Cirrus galaxies at z =1-2 could give similar
250-500$\mu$m colours to starbursts at z = 4-6.  However it would not be be possible for them to be detected
at 500 and 350 $\mu$m, but not at 24$\mu$, 3.6$\mu$m and in the r-band.  So they could not contribute to the
unidentified sources.  There can be no cirrus galaxies at z$>$2 in this sample because their infrared
luminosity would exceed the maximum possible for an optical galaxy.  Thus it is reasonable to use only
starburst templates in estimating the redshifts of unidentified sources.

\begin{figure*}
\includegraphics[width=7cm]{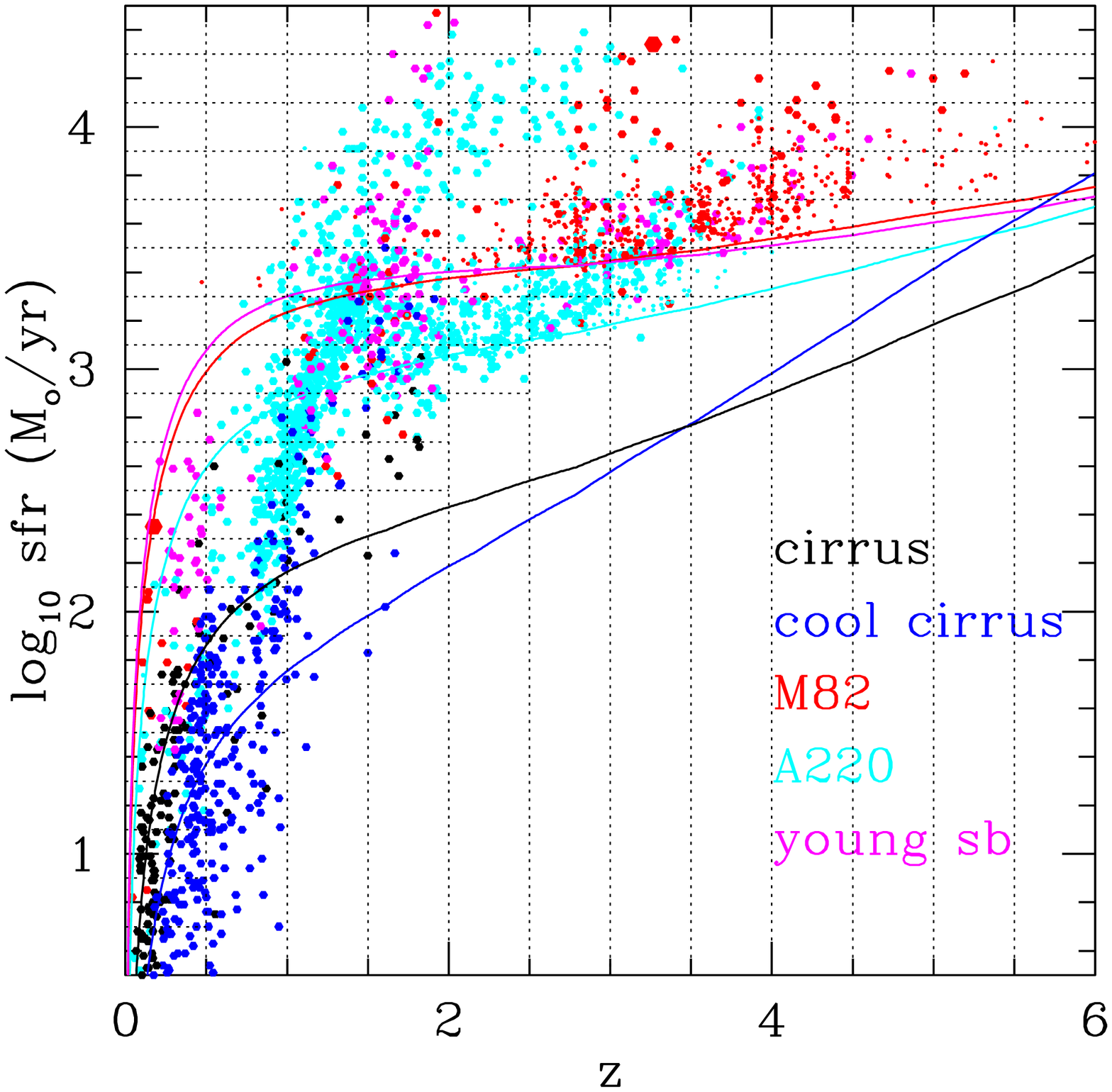}
\includegraphics[width=7cm]{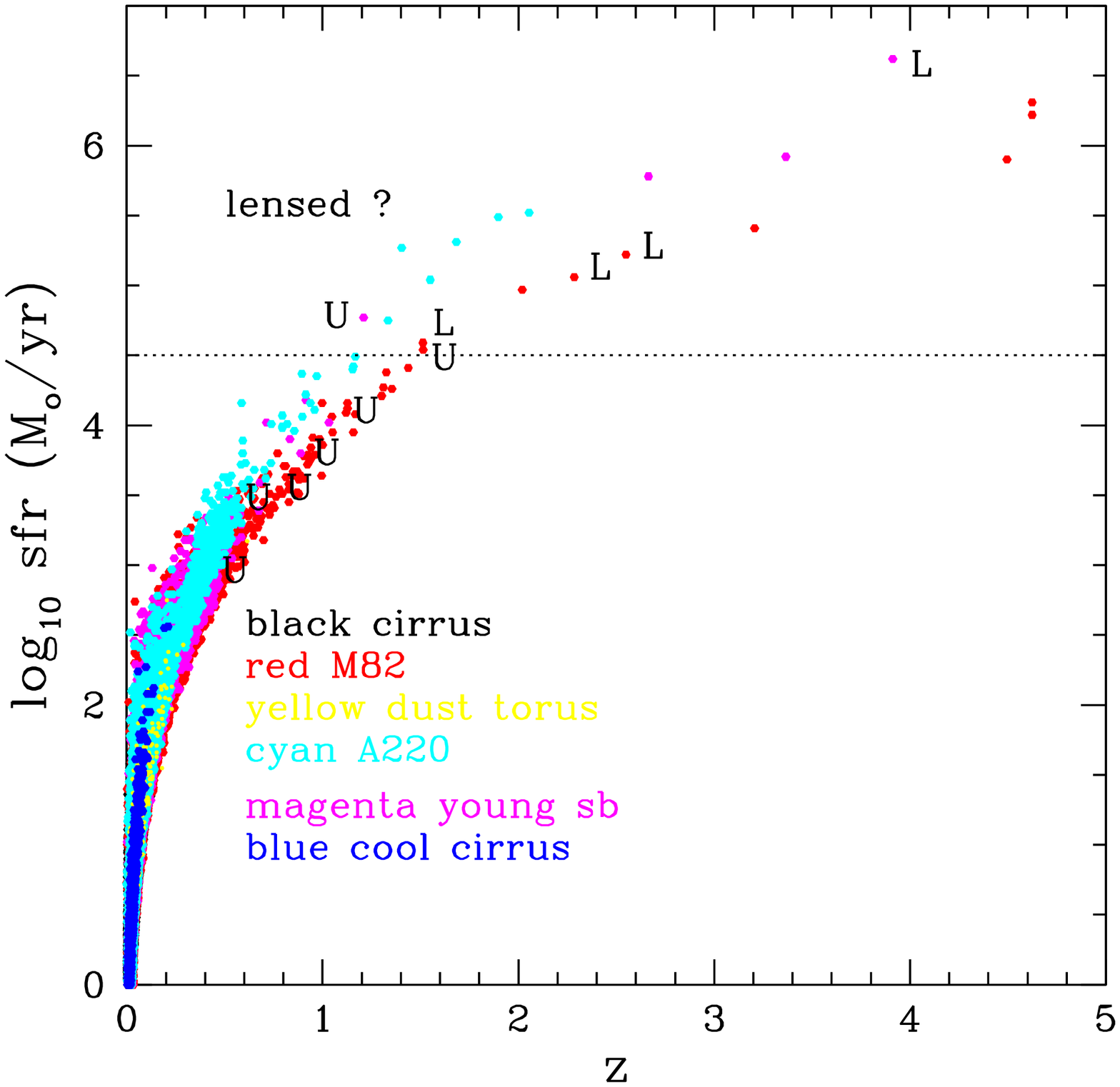}
\caption{L: Star formation rate versus redshift for HerMES Lockman+XMM+ES1 galaxies, with 500 $\mu$m selection limits
for each template type.  
R: Star-formation rate versus redshift for 60,303 IRAS RIFSCz galaxies (Wang et al 2014b).
Known lenses are indicated by L and cases known to be unlensed indicated by U (Farrah et al 2002). 
}
\end{figure*}

\begin{figure}
\includegraphics[width=7cm]{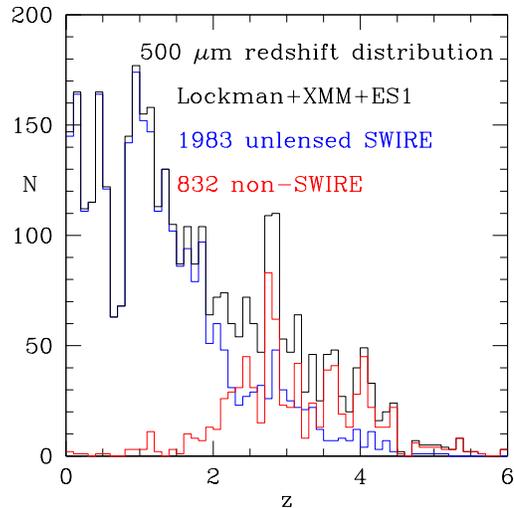}
\caption{Redshift distribution for HerMES-Lockman+XMM+ES1 500 $\mu$m sample.  Blue: SWIRE associated, unlensed;
red:SWIRE unassociated; black: total.
}
\end{figure}

It is striking that star-formation rates extend up to 20,000 $M_{\odot}{\rm yr}$.  Such enormous
star-formation rates can presumably last for no more than $10^7$ yrs, compared with 
$\sim 10^8$ yrs for local starbursts (Genzel et al 1998, Rowan-Robinson 2000). 
Such high star-formation rates have been seen before in the IRAS hyperluminous galaxies (Rowan-Robinson 2000, Rowan-Robinson
and Wang 2010).  Figure 2R shows the star-formation rate versus redshift for the RIFSCz 
catalogue (Wang et al 2014b), with known lensed sources indicated by L, and sources
for which Hubble Space Telescope imaging (Farrah et al 2002) shows no sign of lensing indicated by U.  
The boundary between lensed and unlensed sources appears to fall at a star-formation rate of $\sim$30,000
$M_{\odot}/yr$.  In the case of the all-sky RIFSCz, selected at 60 $\mu$m, we see rare, exceptionally
strongly lensed galaxies, in which the optical emission is also from the lensed galaxy. 
Typically the surface-density of such objects is 0.001 $sq deg^{-1}$.  For the HerMES
500 $\mu$m-selected candidate lensed galaxies, the optical emission is from the lensing galaxy and the surface
density of candidate lenses is 10,000 times greater, $\sim 10 sq deg^{-1}$.

\section{Reliability of z $>$ 4 photometric redshifts}

The reliability of SWIRE photometric redshifts has been discussed by Rowan-Robinson et al (2013).
Provided at least 5 photometric bands are available, photometric redshifts have an rms accuracy
better than 5$\%$ and a catastrophic outlier rate $< 7\%$ out to z = 1.5. Beyond this redshift we have
few spectroscopic redshifts available and the plausibility of redshifts can only be tested through
SED plotting. The 24 and 250-500 $\mu$m fluxes can be helpful here.
Figures 4 and 5 show SEDs of Lockman, XMM-LSS and ES1 500 $\mu$m sources with photometric redshifts $>$ 4.  Most of 
these look plausible. 160.02754+58.22380, the bottom source in Fig 3R, has a blackbody spectrum in the optical and 
is therefore a star.  The submillimetre source has been transferred to the unidentified category. Of the other 26
objects one (162.68120+57.55606) has been fitted with a QSO template at optical-nir wavelengths, but several
others could also be fitted with a QSO template (shown as dotted loci).

\begin{figure*}
\includegraphics[width=7cm]{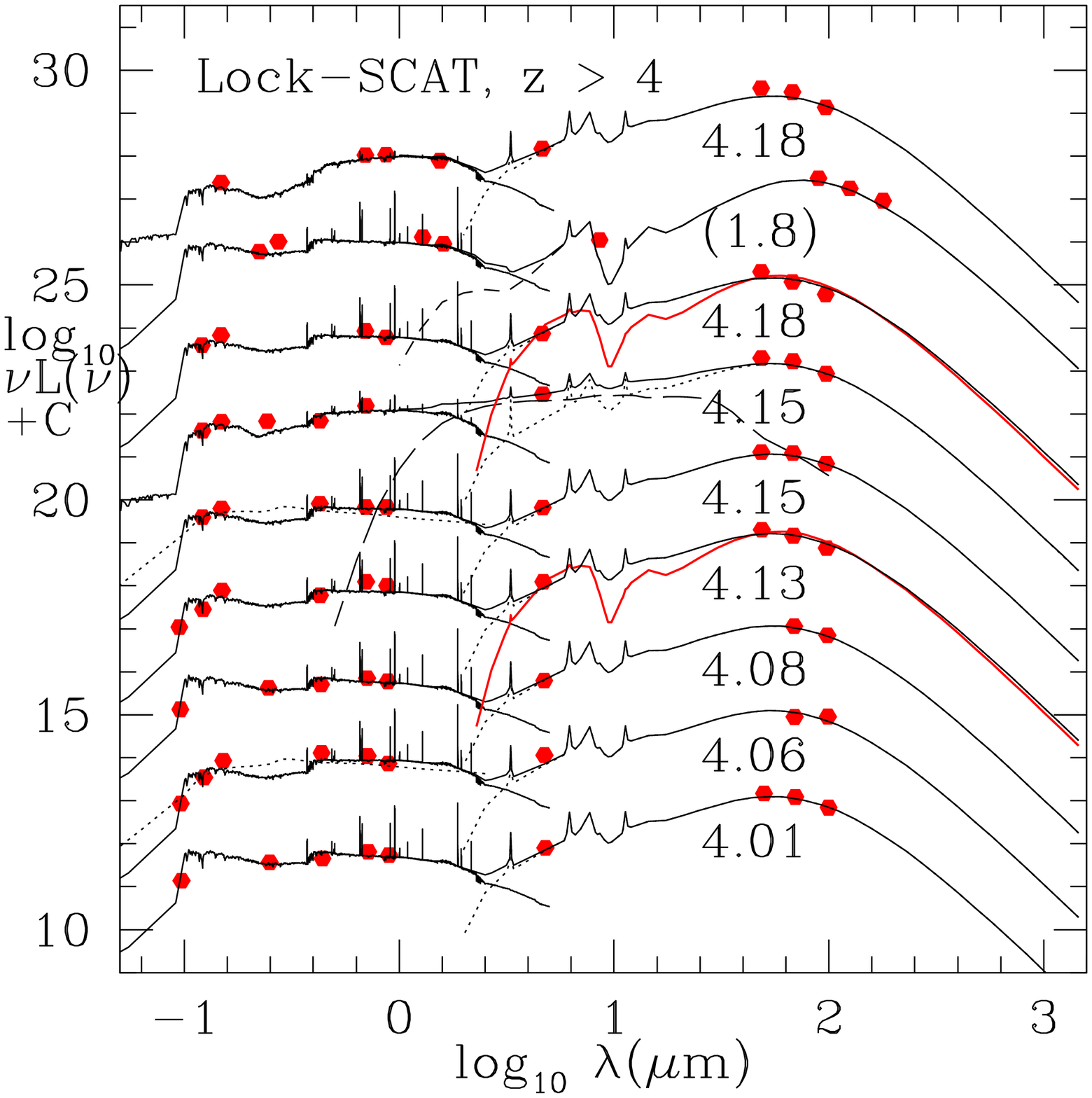}
\includegraphics[width=7cm]{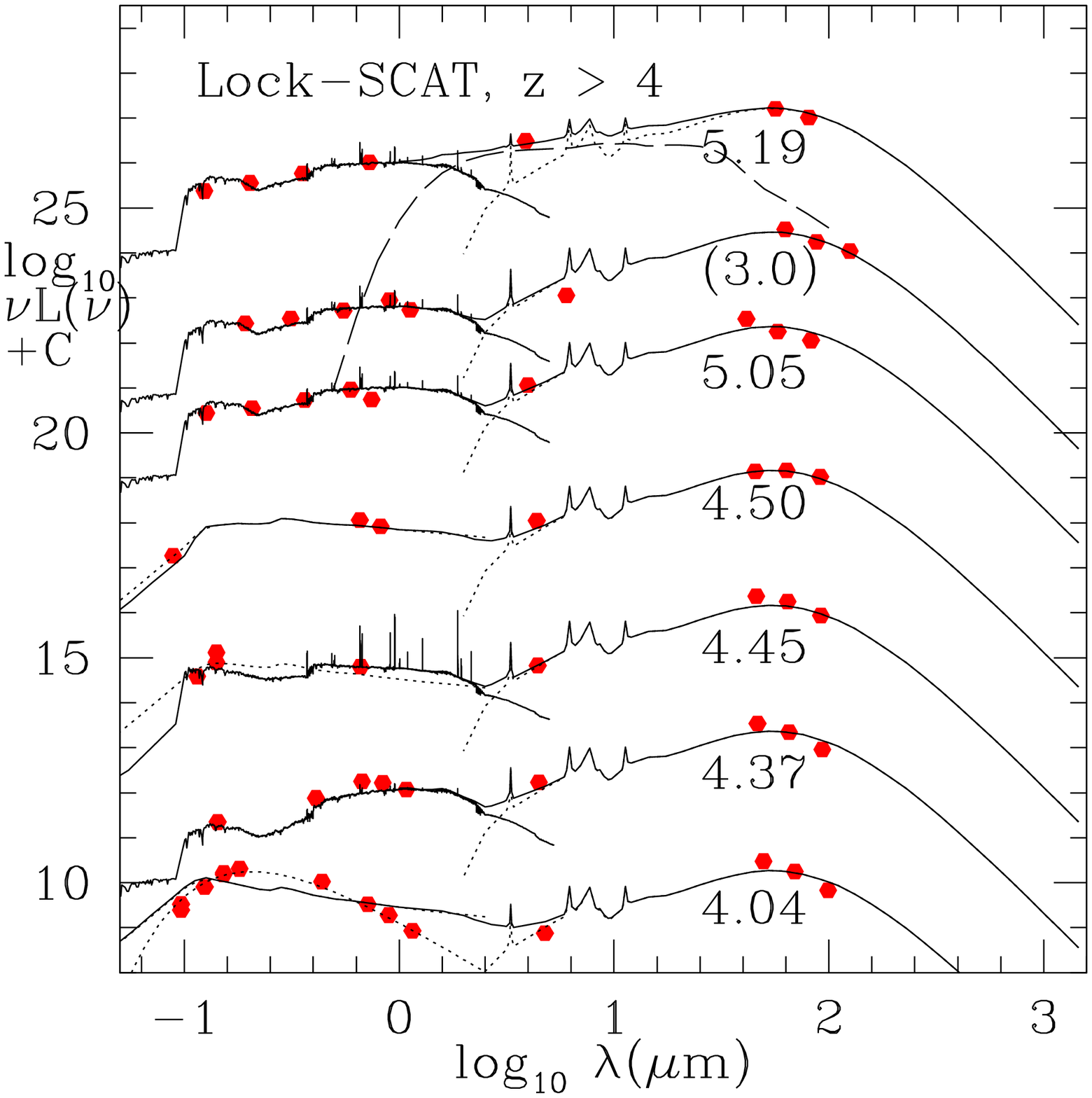}
\caption{
SEDs of Lockman-SPIRE 500 $\mu$m sources with photometric redshifts $>$ 4, labelled with the redshift. 
Lower redshift aliases ($z_{comb}$) are shown with redshift bracketed, above SED for the higher redshift alias. Dotted loci: M82 starburst, dashed loci: Arp 220
starburst, long-dashed loci: AGN dust torus, red loci: young starburst template.
}
\end{figure*}

\begin{figure*}
\includegraphics[width=7cm]{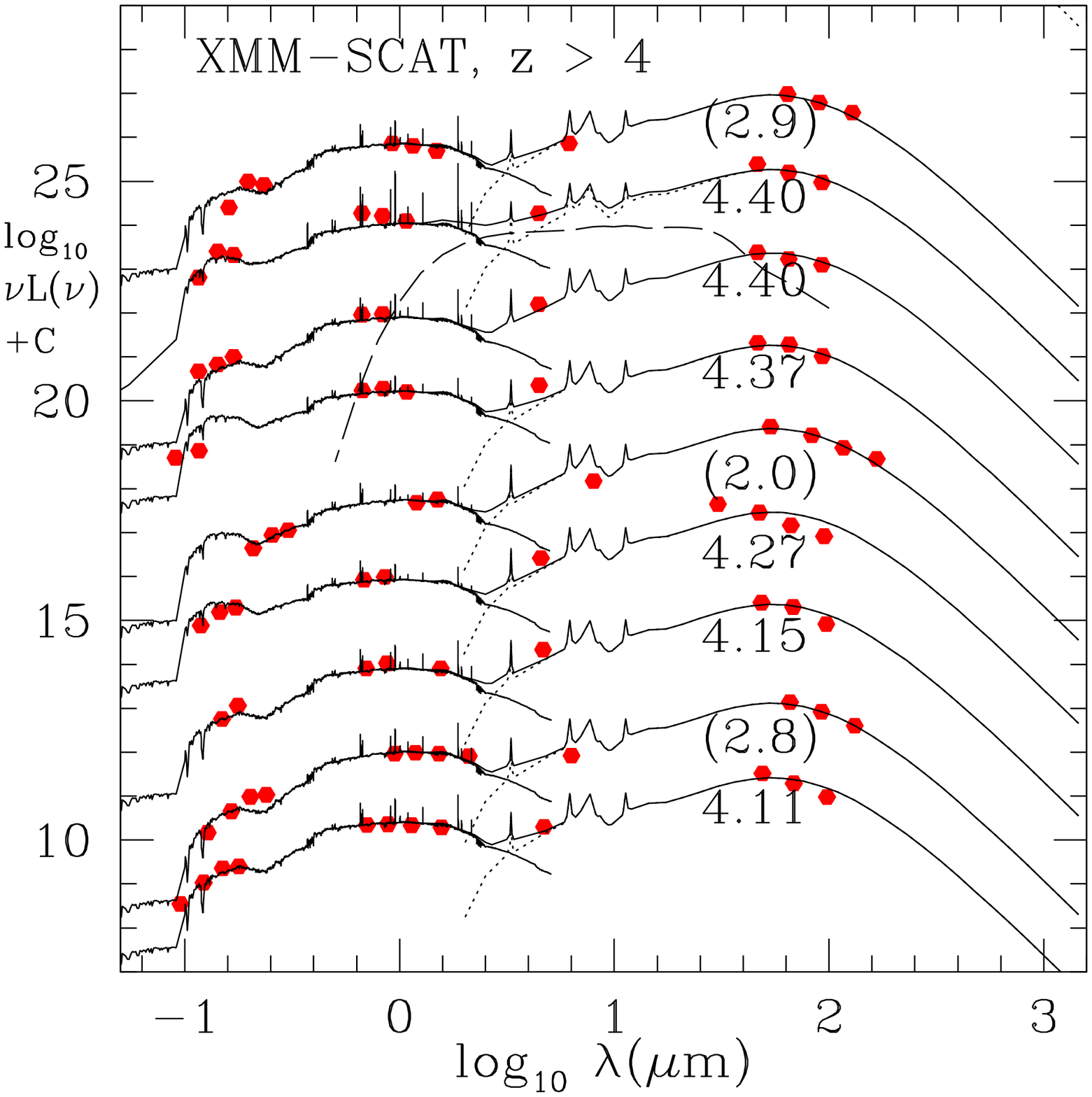}
\includegraphics[width=7cm]{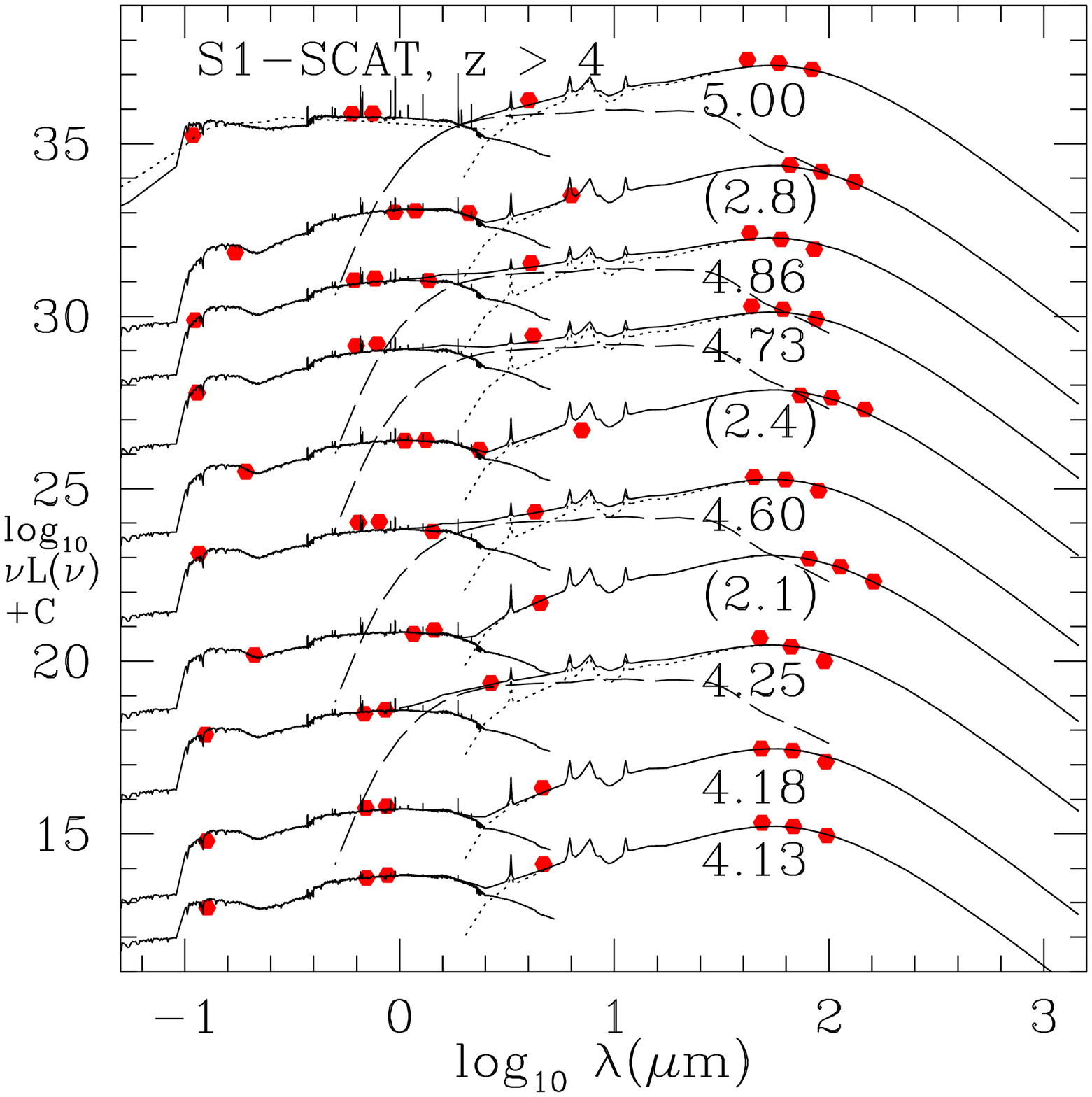}
\caption{
SEDs of SPIRE 500 $\mu$m sources with photometric redshifts $>$ 4 in XMM-LSS (L) and ES1 (R). Notation as for Fig 4.
}
\end{figure*}

We can make an independent estimate of the redshift from the submillimeter (250-500 $\mu$m data), $z_{subm}$,
using the best fit M82, Arp220 or young starburst templates.  This is valuable in selecting between $z_{phot}$
aliases.  For all galaxies with $z_{phot} > 2$ we have estimated $z_{subm}$.  Figure 6L shows $z_{subm}$ versus
$z_{phot}$ for Lockman, XMM-LSS and S1 galaxies. We estimate the uncertainty in $z_{subm}$ as $\pm 30\%$ of (1+z)
(see discussion of $\chi^2$ distributions below).  We have also summed $\chi^2_{phot}$ and
$\chi^2_{subm}$ and estimated $z_{comb}$, where the combined $\chi^2$ is a minimum.
Figure 6R shows a plot of $z_{comb}$ versus $z_{phot}$. Generally $z_{comb} \sim z_{phot}$, with an rms
uncertainty of $\pm 10\%$ in (1+z), showing that the submillimetre data are consistent with $z_{phot}$.
However there are 27 catastrophic outliers (out of 388 galaxies) where (1+$z_{comb}$) differs from (1+$z_{phot}$)
by more than 30$\%$ from (1+$z_{phot}$). 12 of these outliers had $z_{phot} > 4$.  For these 27 outliers we have adopted $z_{comb}$ rather than $z_{phot}$
in the subsequent analysis.  For $z < 2$ it is not possible to make a meaningful estimate of $z_{subm}$
because of the presence of cirrus components, which alias catastrophically with Arp 220 starbursts. 

Figure 7 shows $\chi^2_{phot}$, $\chi^2_{subm}$ and the sum of these for the $z_{phot} > 4$
galaxies in Lockman. The $\chi^2_{subm}$ distributions demonstrate that the uncertainty in $z_{subm}$ is
$\sim\pm 30 \%$ of (1+z), due to the different redshifts generated by the different templates seen in
luminous starbursts.
For galaxies where $z_{comb}$ differs significantly from $z_{phot}$ we have shown
SED fits for both $z_{phot}$ and $z_{comb}$ (labelled with redshifts in brackets) in Fig 4 and 5. 

We have tested the reliability of our $z_{subm}$ estimates by applying them to 28 Herschel sources with
measured spectroscopic redshifts (Cox et al 2011, Lupu et al 2012, Combes et al 2012, Riechers et al 2013, 
Wardlow et al 2013, Dowell et al 2014, Rowan-Robinson et al 2013).  Figure 6R shows $z_{subm}$ versus $z_{spec}$ for these galaxies.
The rms error in $(1+z_{subm})/(1+z_{spec})$ was found to be 21$\%$. 

Table 3 lists all the objects with photometric redshift $>$ 4, together with $z_{subm}$ and $z_{comb}$.  
In the end the reliability of these redshifts can only be demonstrated with optical or
submillimetre spectroscopy.  This limitation, that redshifts $>$4 are entirely photometric, applies equally to the
ultraviolet-based studies of Bouwens et al (2012a,b) and Schenker et al (2013).

\begin{table*}
\caption{Galaxies with photometric redshifts $>$ 4}
\begin{tabular}{llllllllllllll}
RA & dec & i & S24 & S250 & S350 & S500 & $z_{phot}$ & type & $\chi^2$ & $n_{bands}$ & $z_{subm}$ & $z_{comb}$ &\\
&&& ($\mu$Jy) & (mJy) & (mJy) & (mJy) &&&& (opt-nir) &&&\\
Lockman &&&&&&&&&&&&&\\
162.84616 & 58.00514 & (g=24.41) & 179.3 & 34.4 & 40.1 & 32.4 & 4.01 & sb & 13.0 & 5 & 3.6 & 4.0 & \\
162.46065 & 58.11701 & 21.96 & 252.5 & -    & 28.4 & 41.4 & 4.06 & sb (QSO?)& 7.9 & 6 & 4.8 & 4.1 & \\
163.90918 & 57.94004 & (g=24.48) & 135.0 & -    & 36.7 & 31.9 & 4.08 & sb & 12.6 & 5 & 3.9 & 4.0 &\\
161.98271 & 58.07477 & 22.10 & 264.4 & 44.2 & 45.3 & 33.6 & 4.13 & sb & 25.4 & 6 & 3.2 & 3.7 & \\
163.51913 & 58.28540 & 22.35 & 138.4 & 28.1 & 37.2 & 30.3 & 4.15 & sb (QSO?) & 1.0 & 5 & 3.8 & 4.1 & \\
164.28366 & 58.43524 & 22.30 & 596.0 & 43.5 & 51.0 & 37.4 & 4.15 & Scd & 60.0 & 5 & 2.5 & 3.8 & \\
164.02647 & 57.07153 & 22.28 & 252.3 & 43.2 & 35.2 & 26.0 & 4.18 & sb & 2.5 & 4 & 2.7 & 1.8 & low-z alias\\
164.52054 & 58.30782 & 23.40 & 306.9 & 81.9 & 92.1 & 58.2 & 4.18 & Sab & 0.03 & 3 & 2.2 & 3.1 & \\
&&&&&&&&&&&&&\\
160.02754 & 58.22380 & 18.74 & 168.2 & 69.5 & 57.5 & 31.4 & 4.04 & QSO & 61.6 & 11 & 2.4 & & star\\
161.89894 & 58.16401 & 23.60 & 315.0 & 66.4 & 59.7 & 35.3 & 4.37 & Sab & 22.7 & 4 & 1.8 & 2.9 & \\
162.42290 & 57.18750 & 22.25 & 121.0 & 43.5 & 46.1 & 32.6 & 4.45 & sb (QSO?) & 5.6 & 4 & 3.2 & 4.4 & \\
162.68120 & 57.55606 & (g=24.35) & 194.9 & 25.5 & 37.0 & 38.3 & 4.50 & QSO & 1.3 & 3 & 4.8 & 4.5 & \\
161.58835 & 59.65826 & 23.71 & 156.0 & 48.1 & 35.1 & 31.8 & 5.05 & Scd & 4.7 & 5 & 3.0 & 3.0 & \\
161.63013 & 59.17688 & 23.94 & 391.4 & -    & 29.6 & 27.0 & 5.19 & Scd & 4.8 & 4 & 3.0 & 4.8 & \\
&&&&&&&&&&&&\\
ES1 &&&&&&&&&&&&&\\
7.98209 & -43.29812 & R=24.87 & 275.8 & 45.6 & 49.7 & 38.8 & 4.13 & Sab & 0.02 & 3 & 2.5 & 3.4 & \\
9.28433 & -44.23750 & R=25.06 & 437.0 & 62.5 & 72.5 & 51.5 & 4.18 & Scd & 0.03 & 3 & 3.4 & 3.5 & \\
9.08571 & -42.59628 & R=24.89 & 271.8 & 95.7 & 76.3 & 41.3 & 4.25 & Scd & 0.85 & 3 & 2.3 & 2.1 & low-z alias \\
9.11142 & -42.84052 & R=24.44 & 354.5 & 37.6 & 45.0 & 30.1 & 4.60 & Scd & 0.03 & 3 & 2.1 & 2.4 & low-z alias\\
8.70199 & -44.48560 & R=25.38 & 423.0 & 31.2 & 36.1 & 27.1 & 4.73 & Scd & 0.01 & 3 & 3.5 & 3.9 & \\
9.30474 & -43.03506 & R=25.19 & 504.5 & 39.6 & 36.3 & 26.3 & 4.86 & Scd & 0.09 & 3 & 2.9 & 2.8 & \\
9.19681 & -44.42382 & R=24.35 & 249.1 & 39.2 & 43.2 & 41.0 & 5.00 & sb (QSO?) & 0.04 & 3 & 4.0 & 4.1 & \\
&&&&&&&&&&&&\\
XMM-LSS &&&&&&&&&&&&&\\
35.73369 & -5.62305 & 23.42 & 422.0 & 73.1 & 60.9 & 42.1 & 4.11 & Sbc & 4.1 & 6 & 2.6 & 2.8 & \\
34.26031 & -4.95556 & 24.95 & 451.0 & 55.2 & 61.6 & 35.6 & 4.15 & Sbc & 3.9 & 4 & 2.0 & 2.2 & low-z alias \\
36.25277 & -5.59534 & 23.94 & 511.7 & 57.9 & 42.0 & 33.6 & 4.27 & Scd & 1.2 & 5 & 2.7 & 2.0 & low-z alias\\
35.73605 & -4.88950 & (g=25.02) & 421.5 & 40.6 & 51.9 & 40.5 & 4.37 & Scd & 11.4 & 6 & 3.7 & 2.6 & \\
36.65871 & -4.14628 & 24.91 & 288.8 & 46.1 & 45.7 & 48.1 & 4.40 & Sbc & 2.1 & 6 & 4.0 & 4.4 & \\
34.53469 & -5.00769 & 23.47 & 342.8 & 47.3 & 42.4 & 35.8 & 4.40 & sb  & 4.0 & 5 & 2.3 & 2.9 & \\
\end{tabular}
\end{table*}

\begin{figure*}
\includegraphics[width=5.0cm]{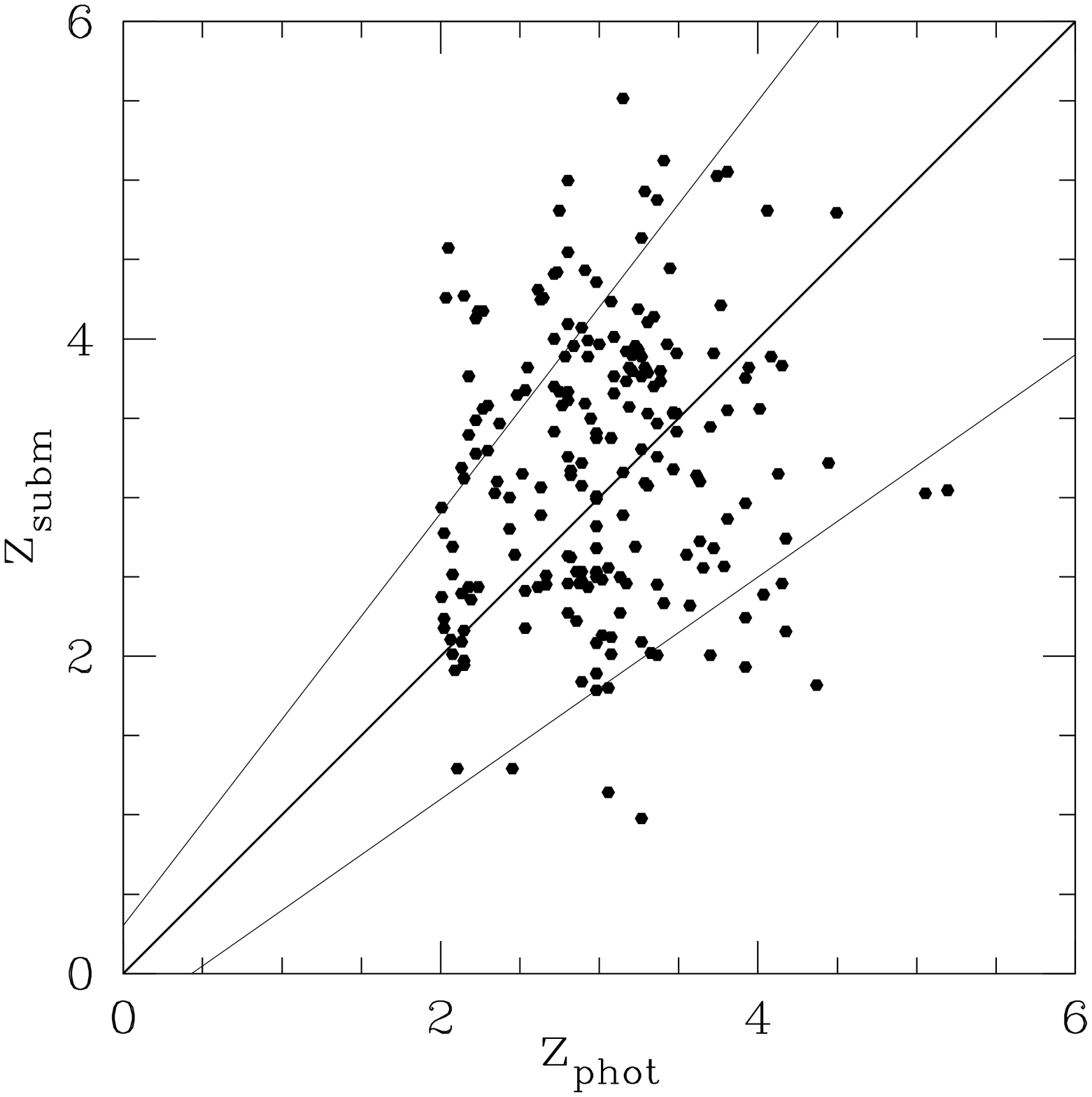}
\includegraphics[width=5.0cm]{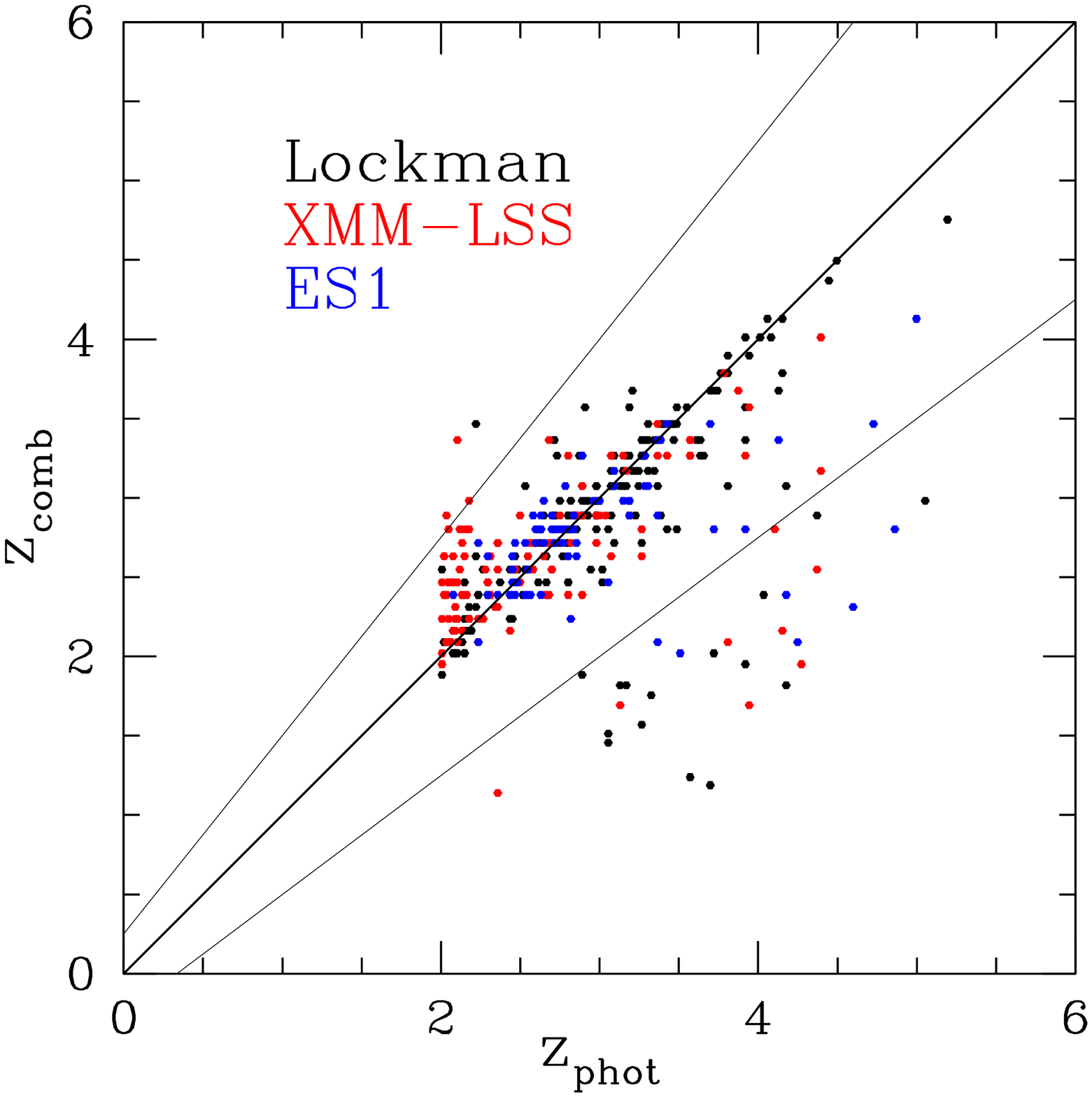}
\includegraphics[width=5.0cm]{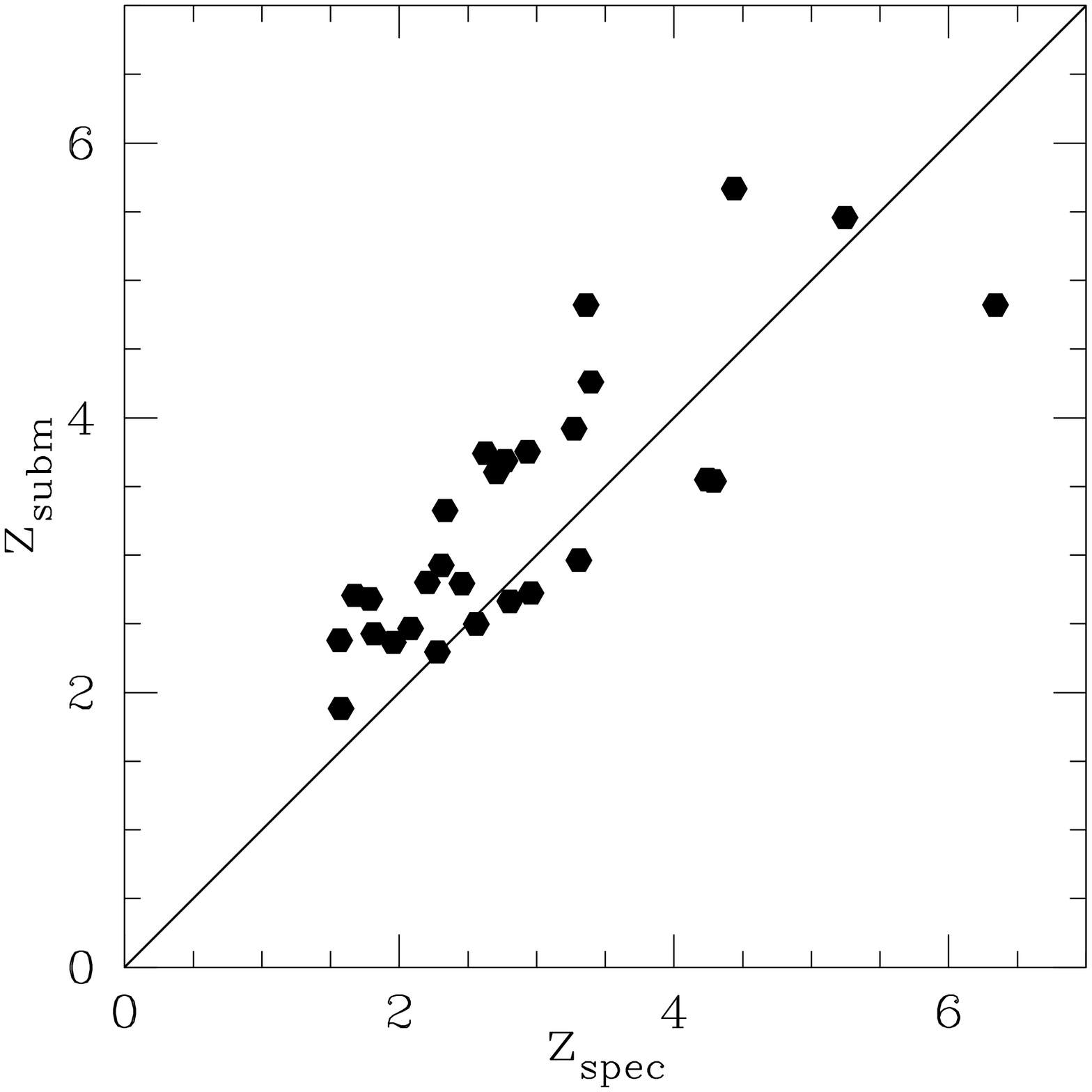}
\caption{L: $z_{subm}$ versus $z_{phot}$ for Lockman, XMM and ES1 galaxies with $z_{phot} > 2$.  The
sloping loci correspond to $\pm 30 \%$ of (1+$z_{phot}$).
C: $z_{comb}$ versus $z_{phot}$ for Lockman, XMM and ES1 galaxies with $z_{phot} > 2$.  Outliers have been assigned redshift $z_{comb}$.  R: $z_{subm}$ versus $z_{spec}$ for Herschel galaxies with spectroscopic redshifts.  Spectroscopic data
from Cox et al 2011, Lupu et al 2012, Combes et al 2012, Riechers et al 2013, Wardlow et al 2013, Dowell et al 2014,
Rowan-Robinson et al 2013.
}
\end{figure*}


\begin{figure*}
\includegraphics[width=5.0cm]{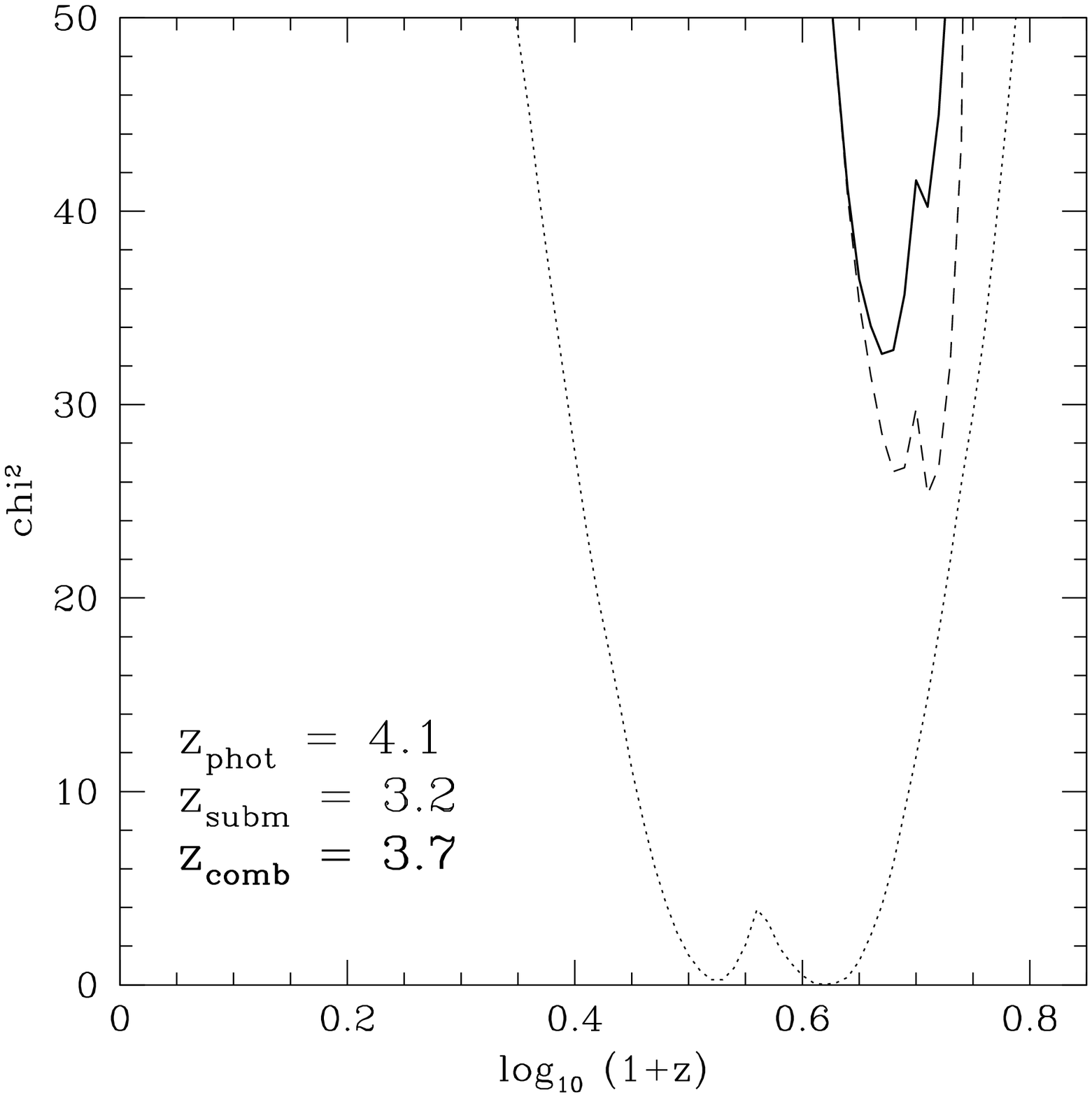}
\includegraphics[width=5.0cm]{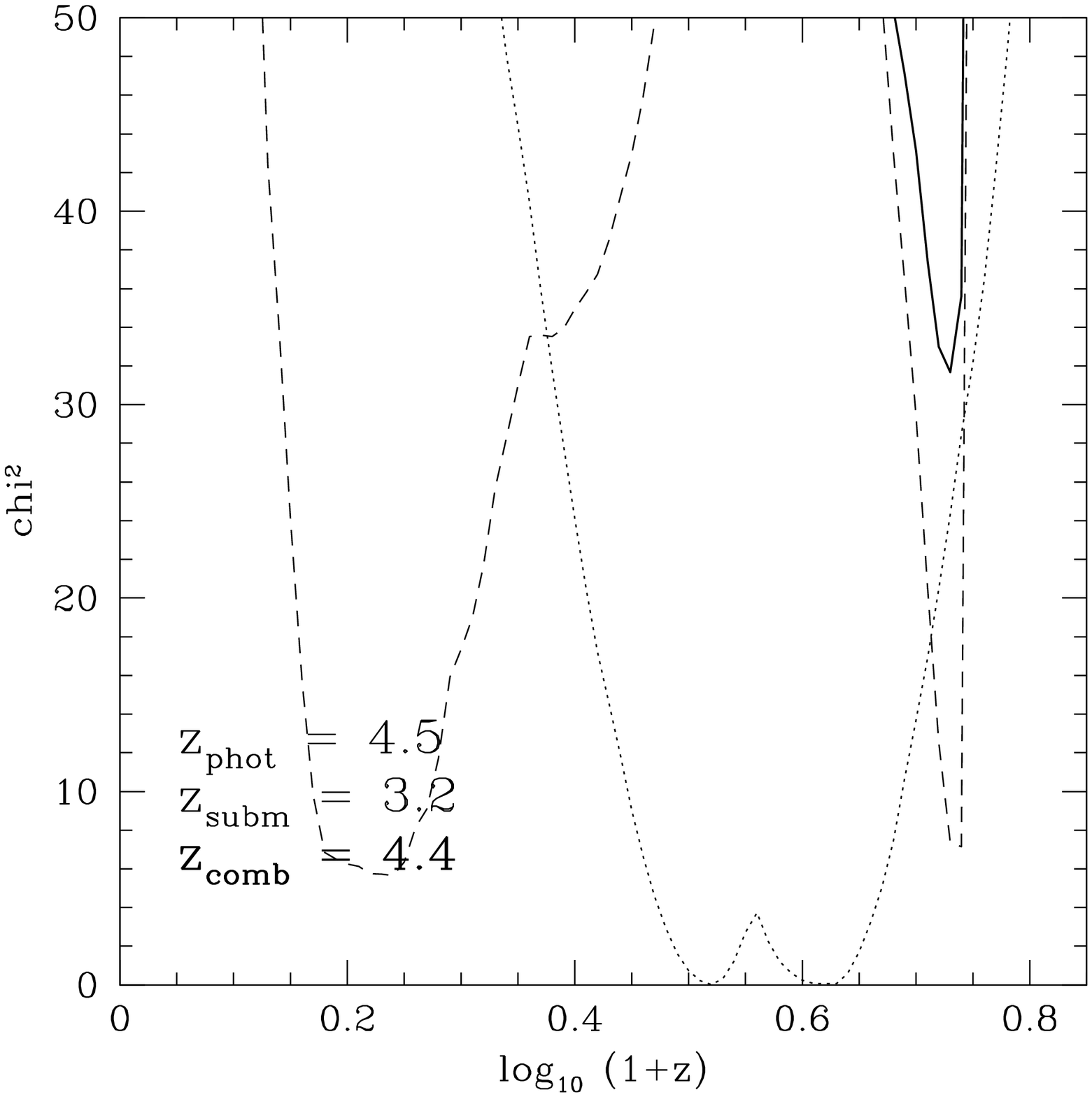}
\includegraphics[width=5.0cm]{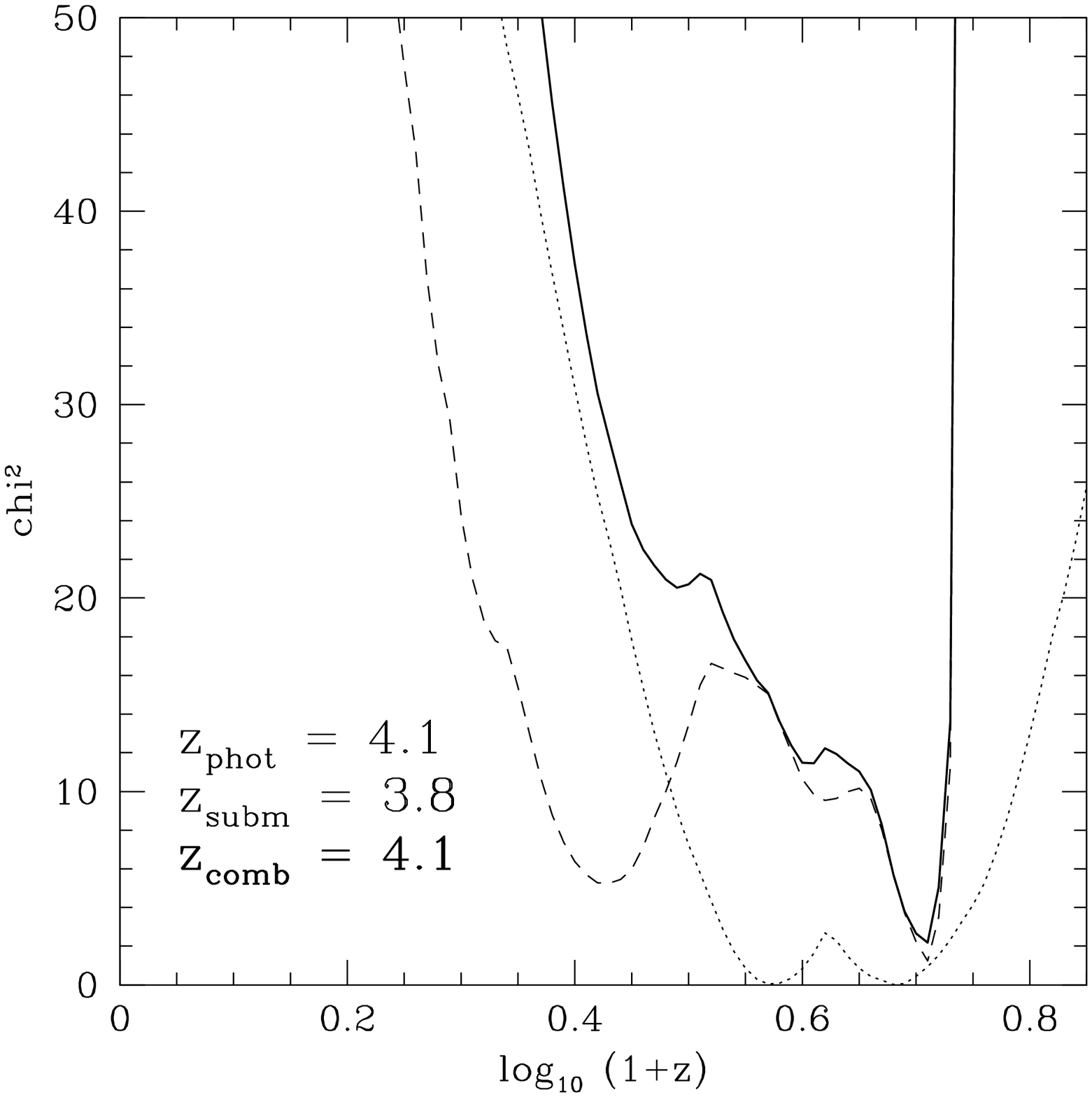}
\includegraphics[width=5.0cm]{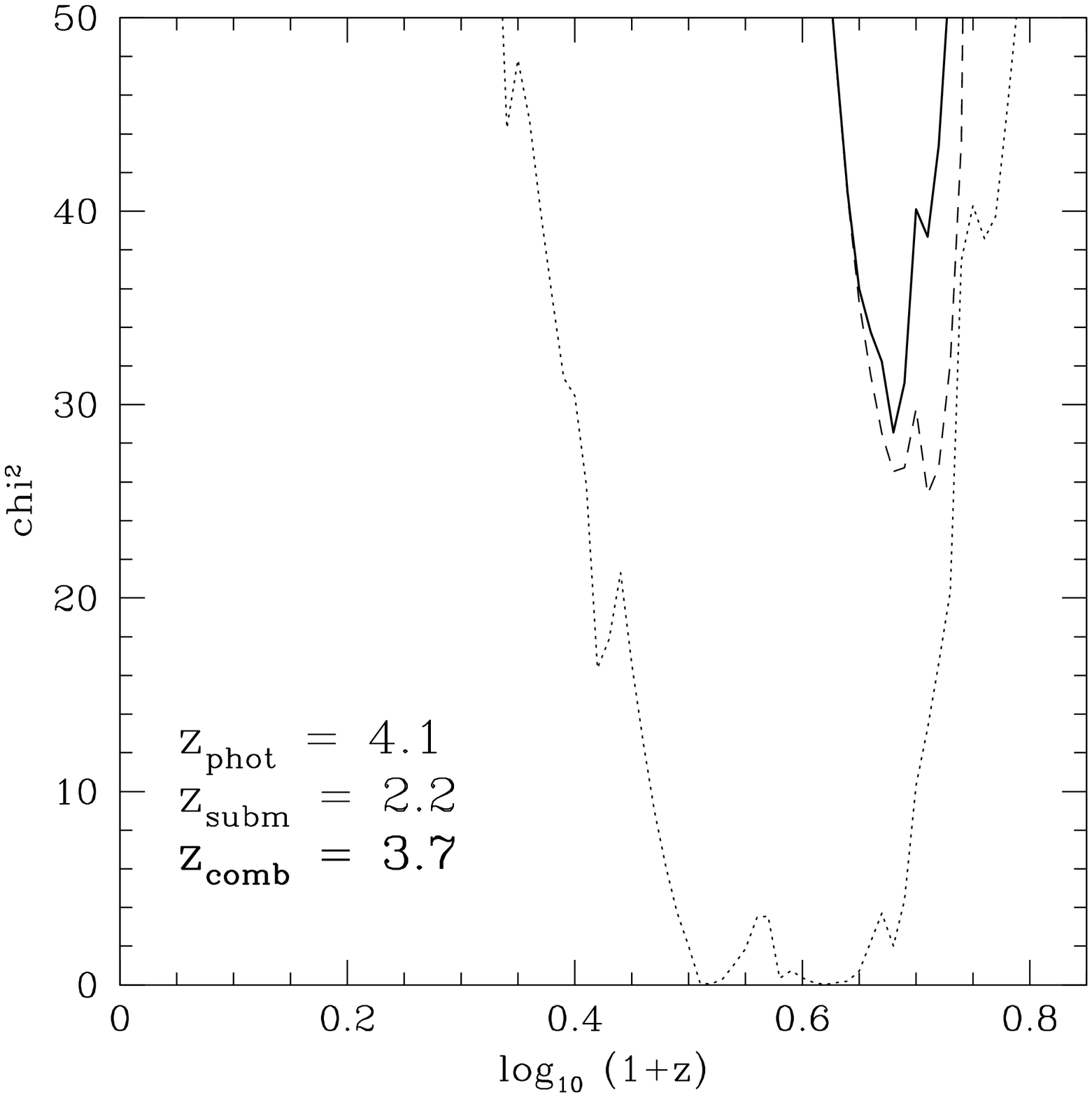}
\includegraphics[width=5.0cm]{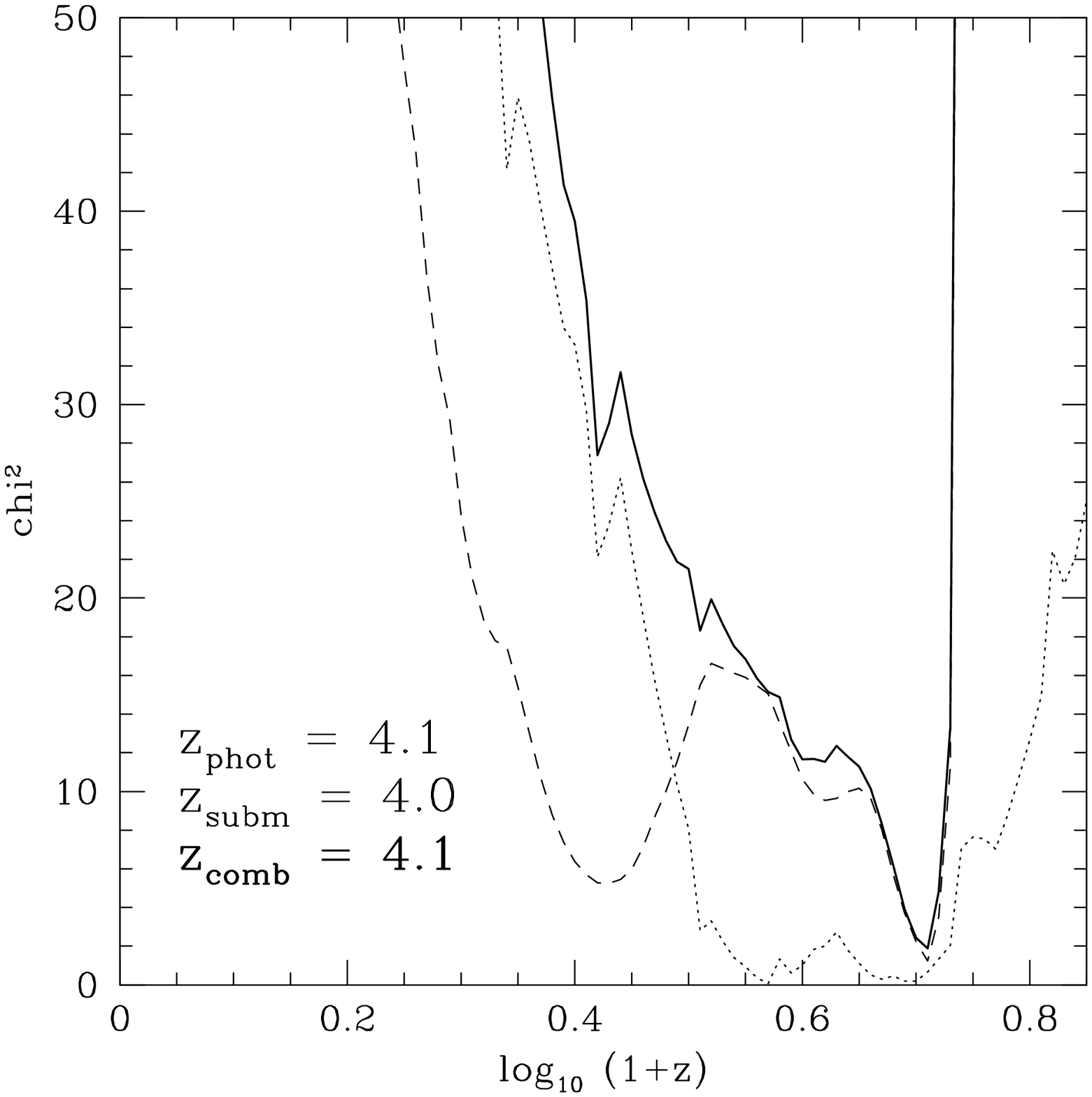}
\includegraphics[width=5.0cm]{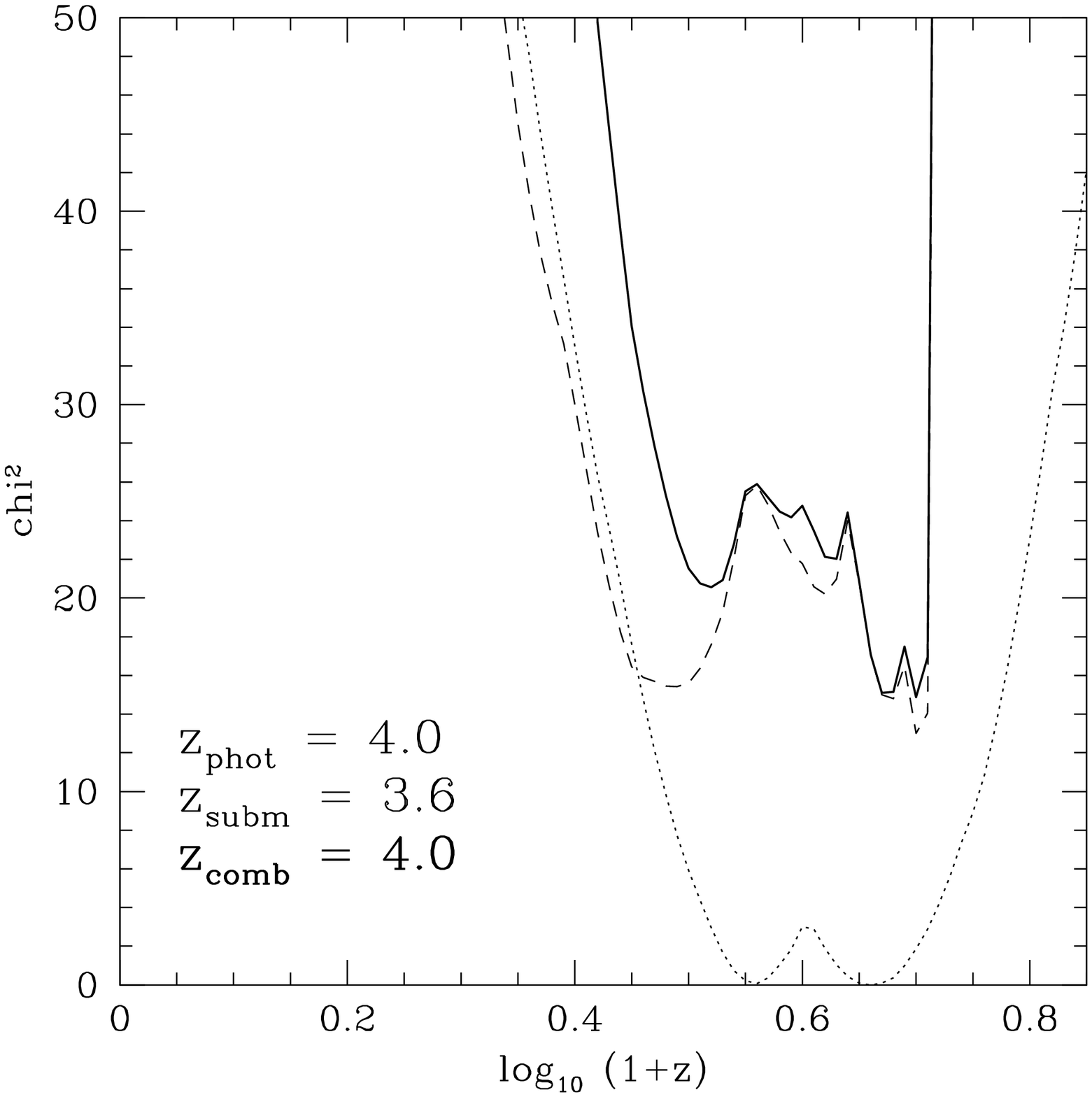}
\includegraphics[width=5.0cm]{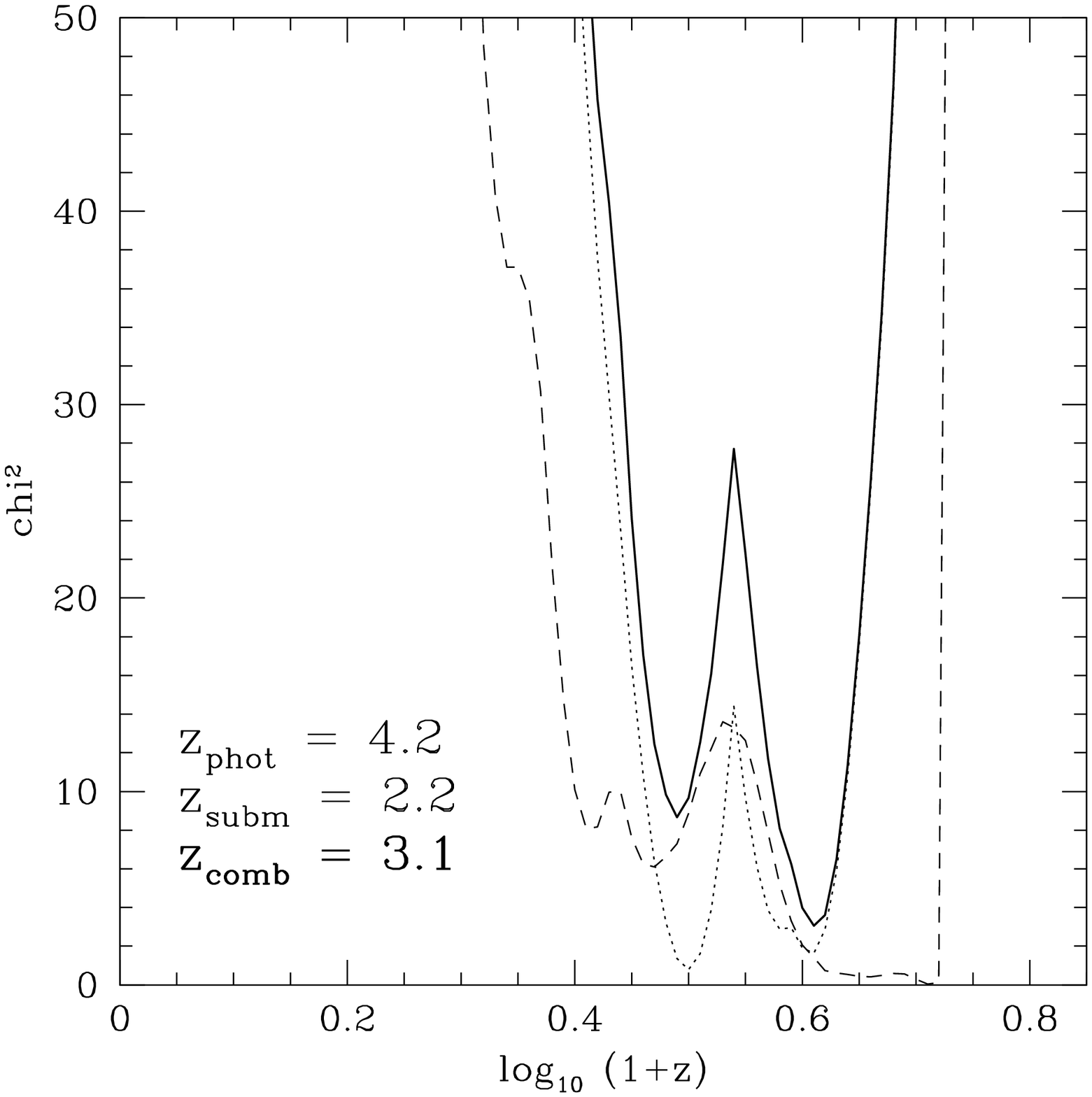}
\includegraphics[width=5.0cm]{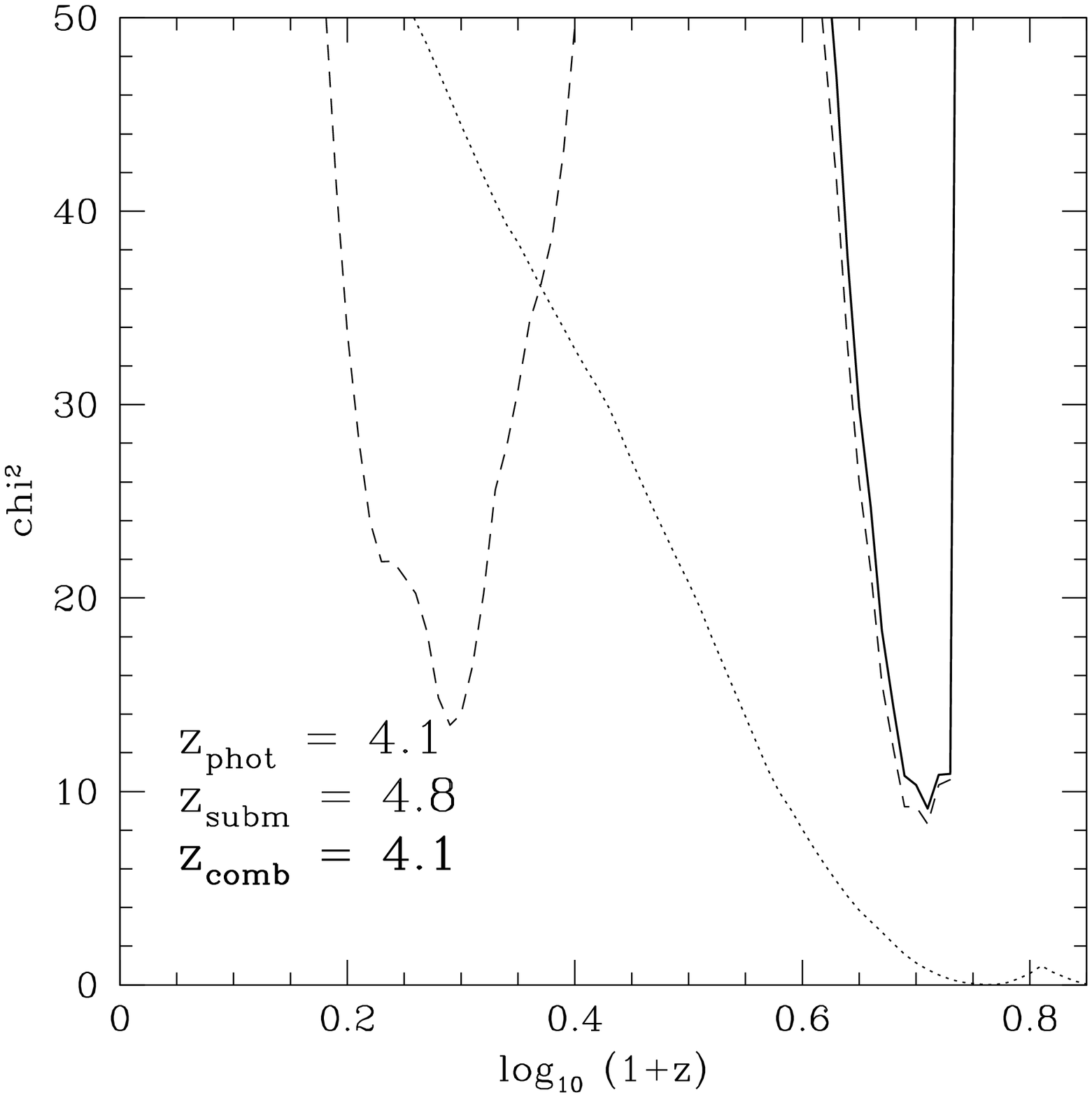}
\includegraphics[width=5.0cm]{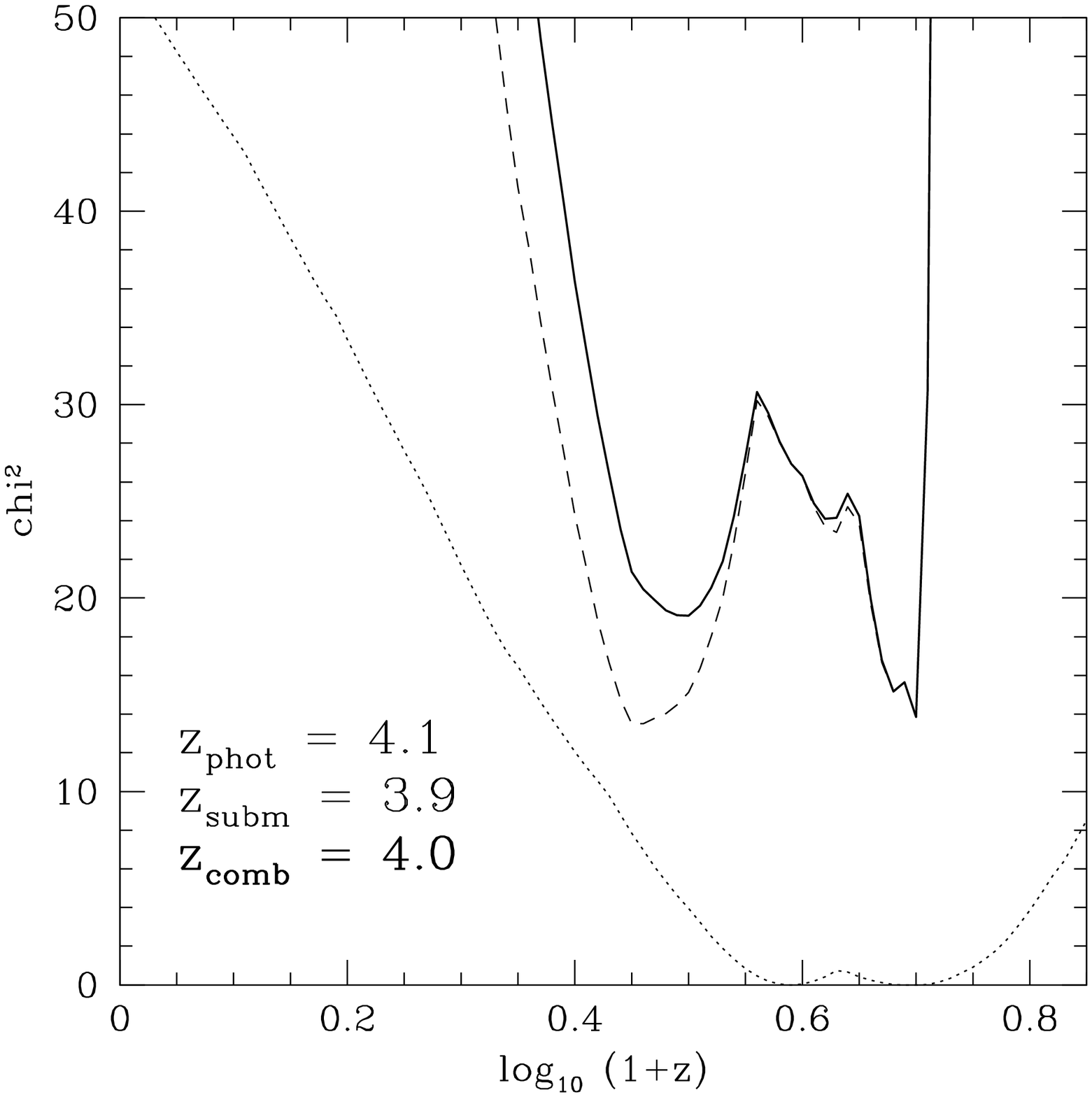}
\includegraphics[width=5.0cm]{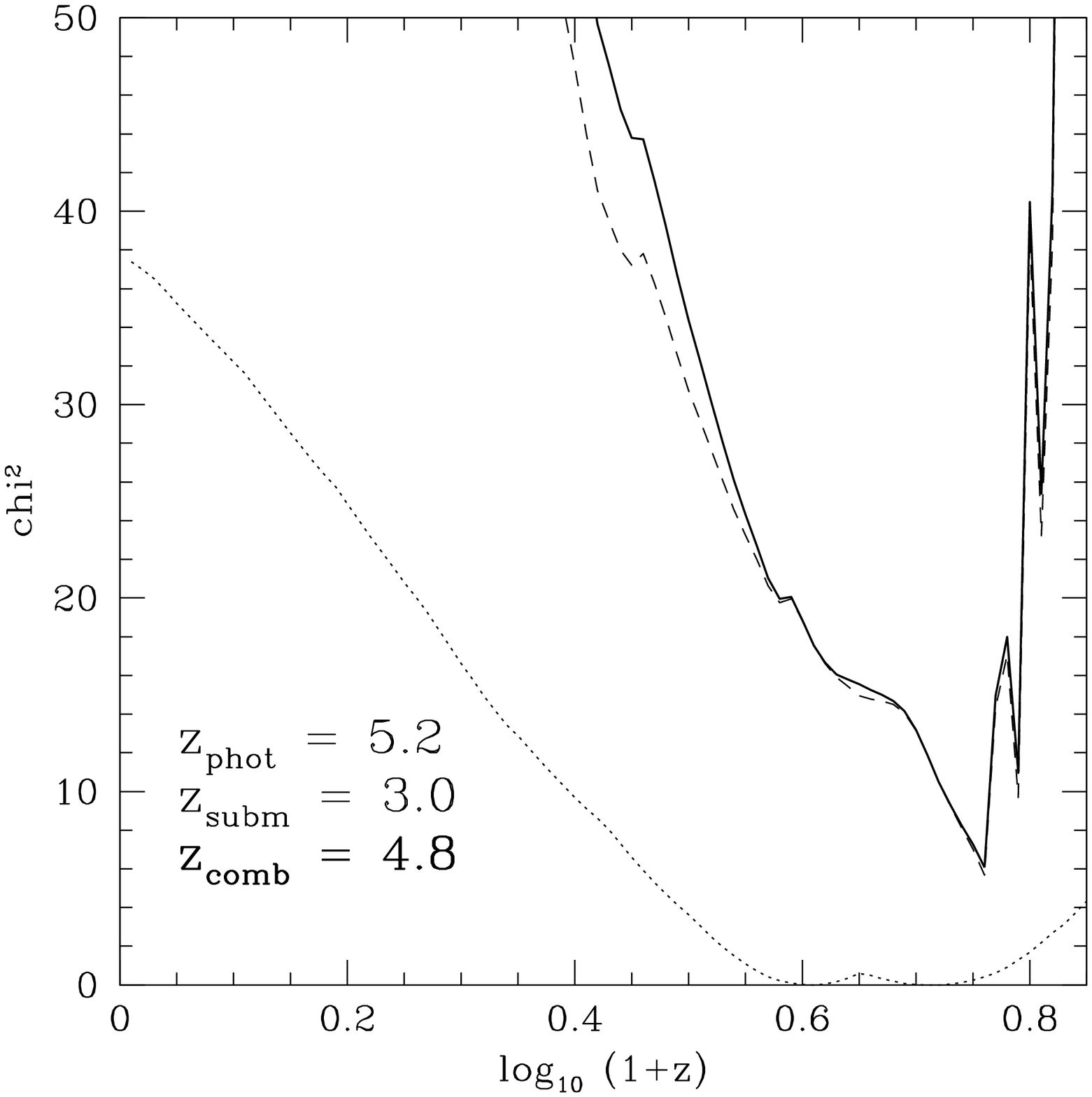}
\includegraphics[width=5.0cm]{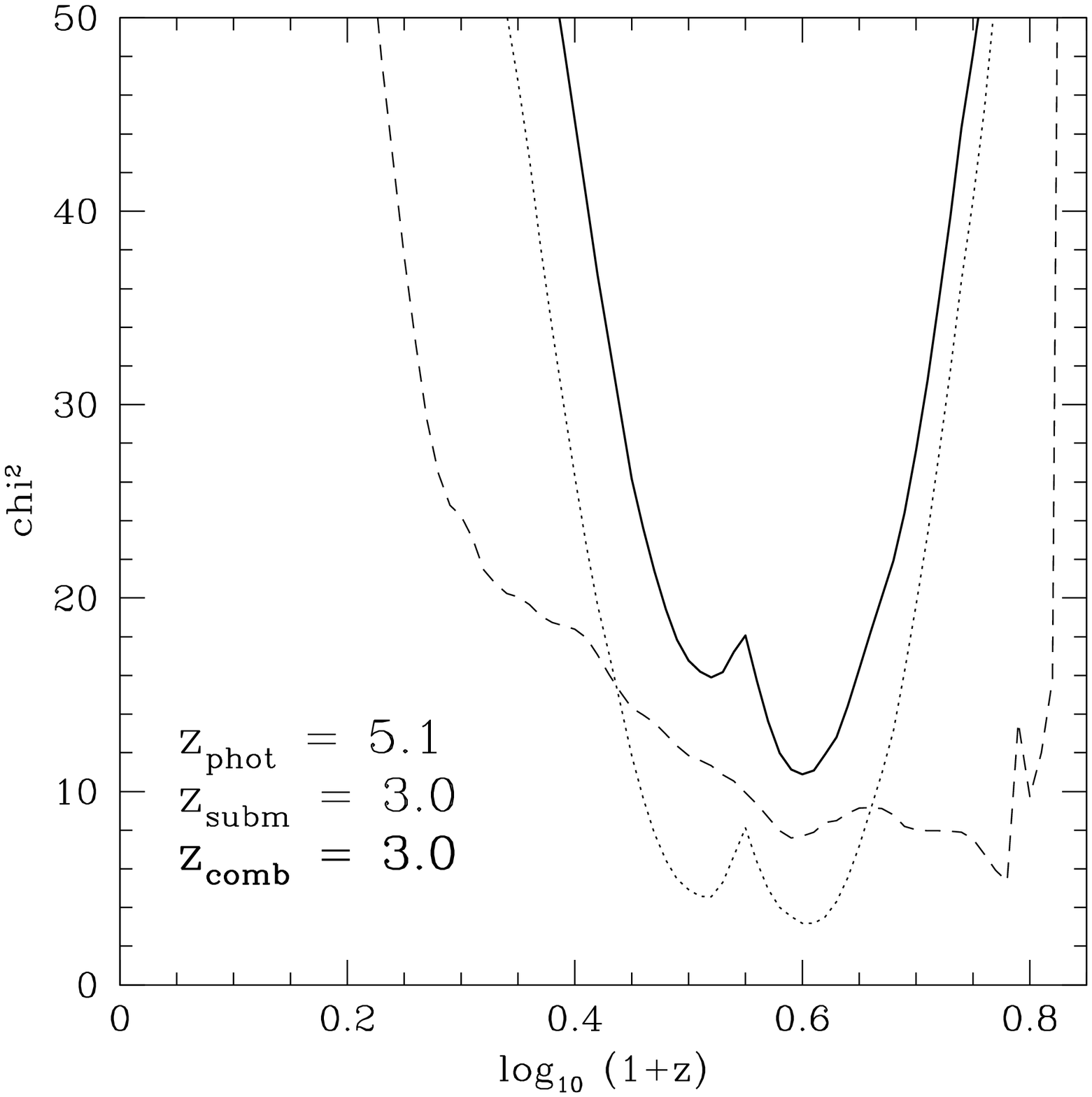}
\caption{$\chi^2$ distributions for Lockman-SPIRE galaxies with $z_{phot} > 4$.  Broken curves: photometric redshift,
dotted curves: submillimetre (250-500 $\mu$m) redshift, solid curves: combined $\chi^2$.
}
\end{figure*}

\section{The star-formation rate function}

We can use the data displayed in Fig 2L (and the corresponding data in XMM-LSS and ES1) to determine the star-formation 
rate function over the redshift range 0-6.  We use bins of 0.5 in z, and 0.2 dex in SFR = $log_{10}(sfr)$.  Ideally we 
would do this
for each template type but our sample is not large enough to do this so we perform this analysis
for the two broad population types, starburst (M82+A220+young starburst) and quiescent
(normal cirrus + cool dust + cold dust).  For each of these two populations we use the most
conservative sensitivity limit locus, the M82 template for starbursts and the normal cirrus template
for quiescent galaxies.

\begin{figure*}
\includegraphics[width=7.0cm]{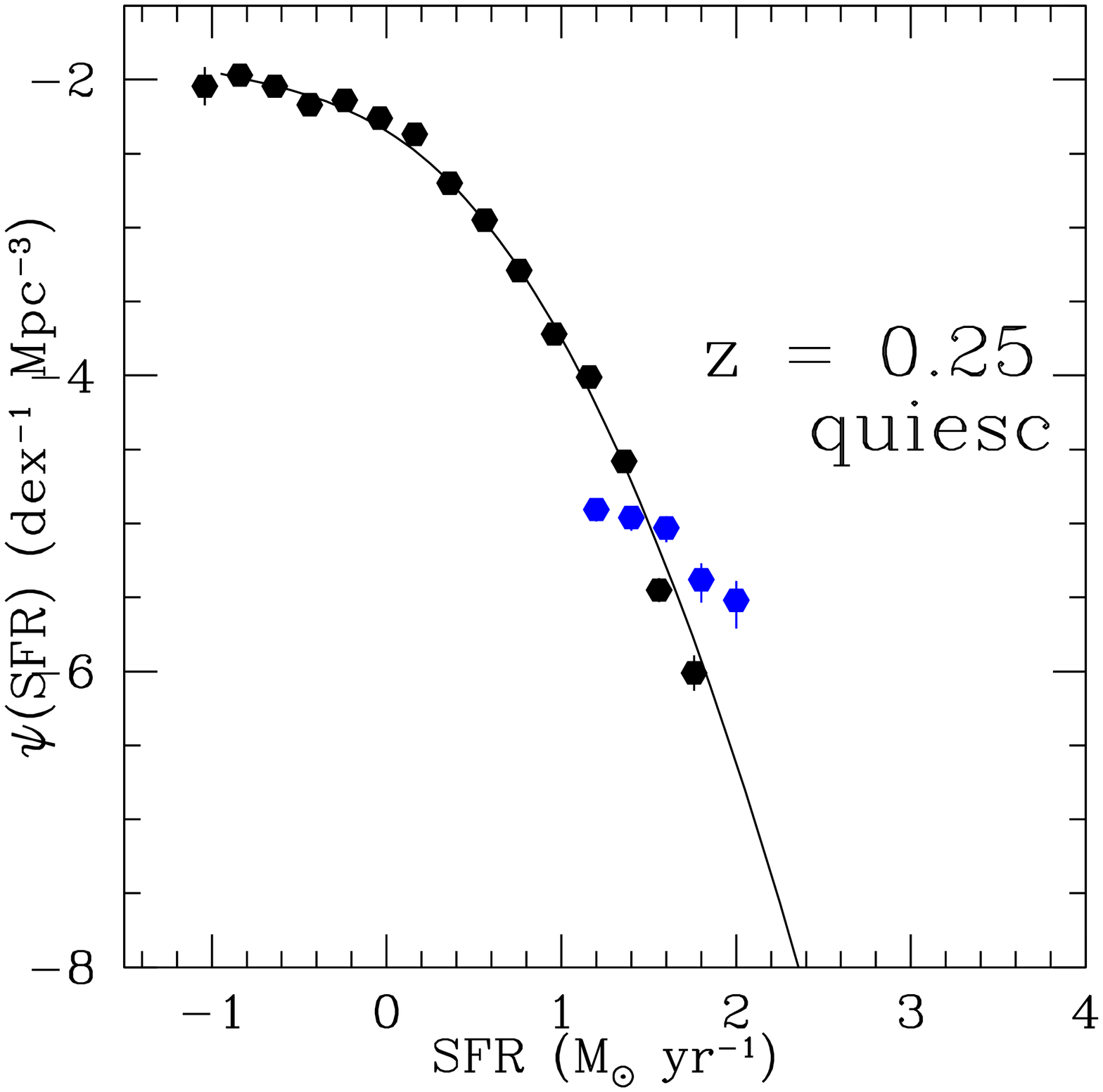}
\includegraphics[width=7.0cm]{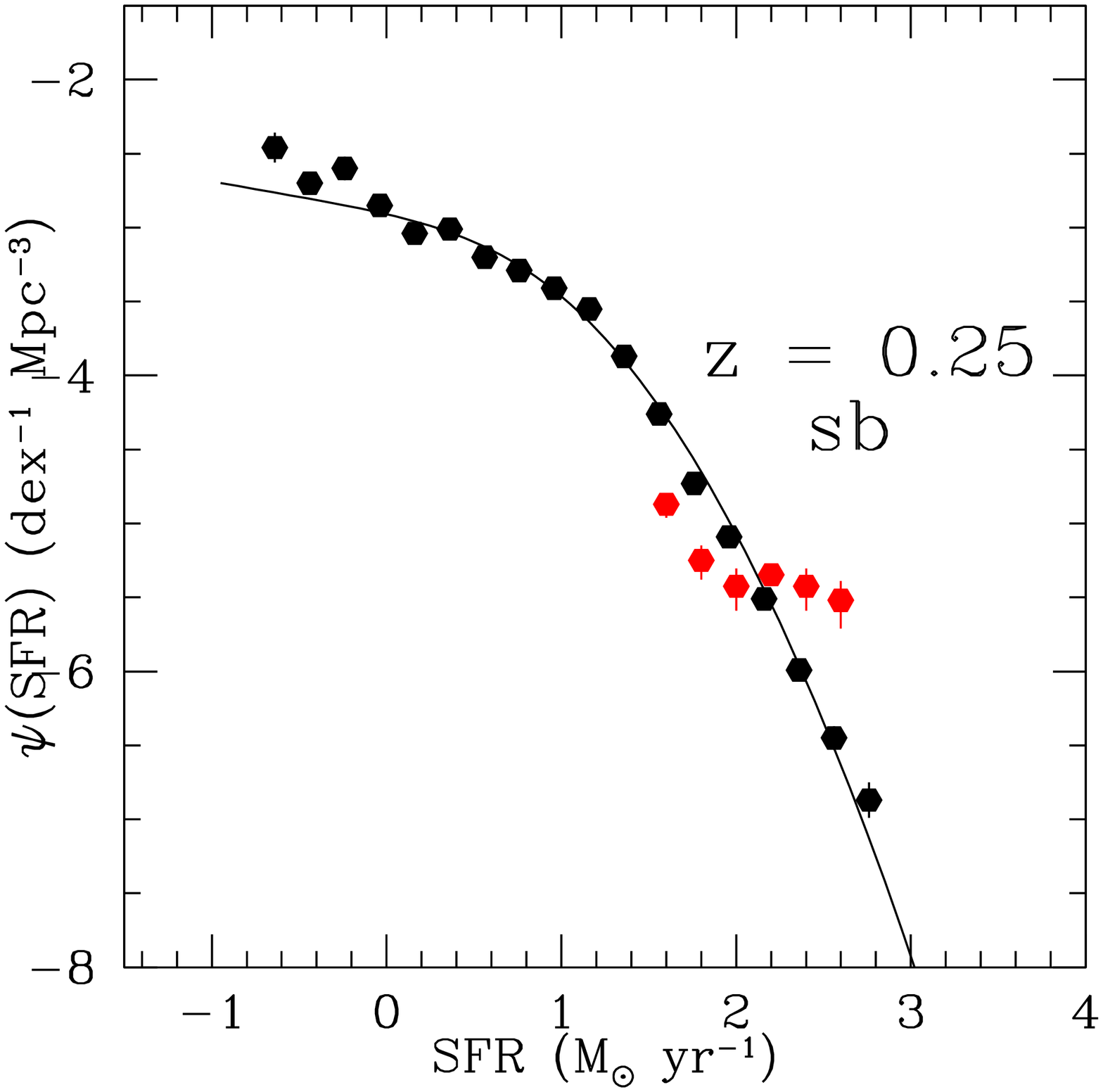}
\caption{
Star formation rate functions for z = 0-0.5.  L: quiescent galaxies (our data in blue), R: starburst
galaxies (our data in red). Black points are adapted from Saunders et al (1990) for ‘cool’ galaxies (L) 
and ‘warm’ galaxies (R). 
}
\end{figure*}

\begin{figure*}
\includegraphics[width=5.0cm]{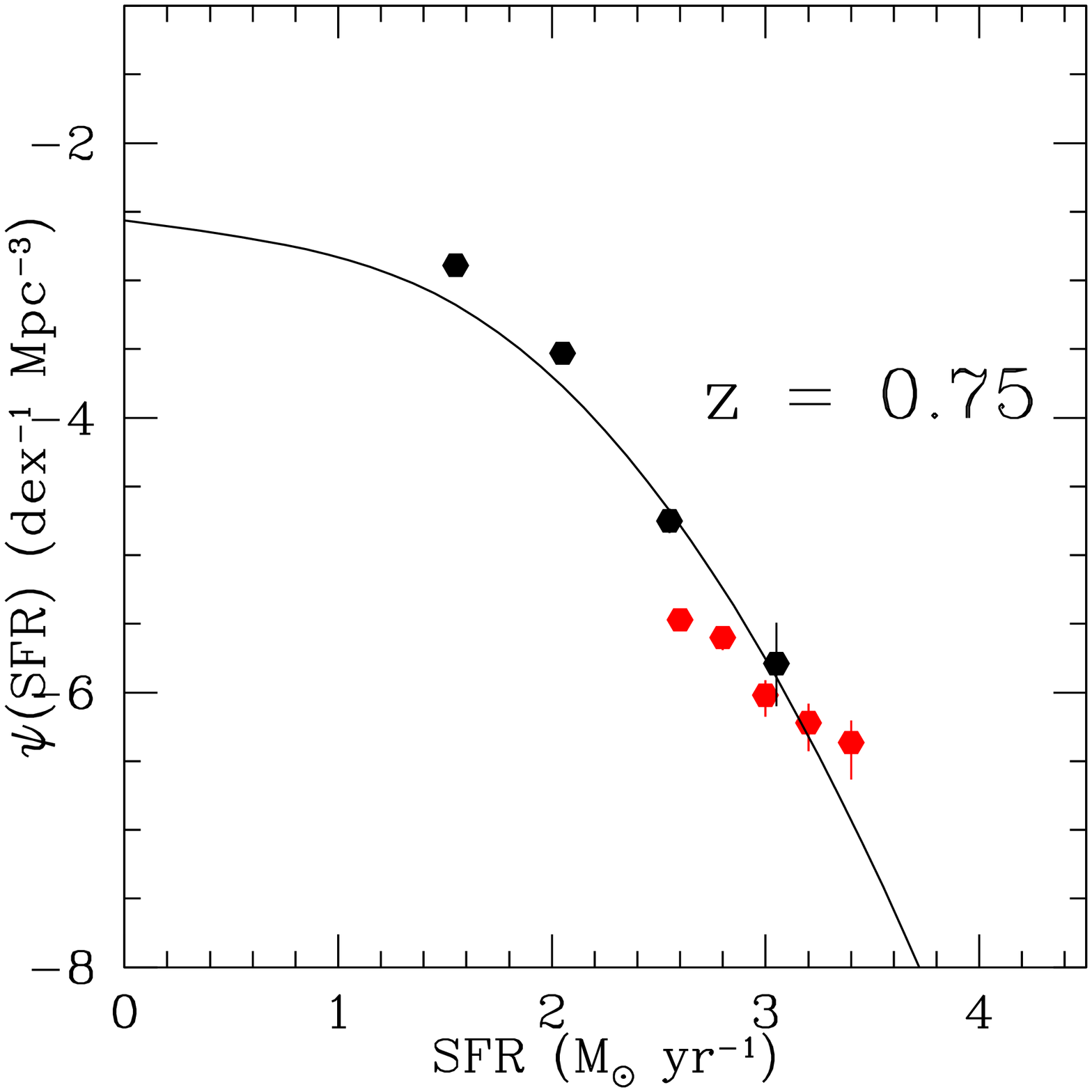}
\includegraphics[width=5.0cm]{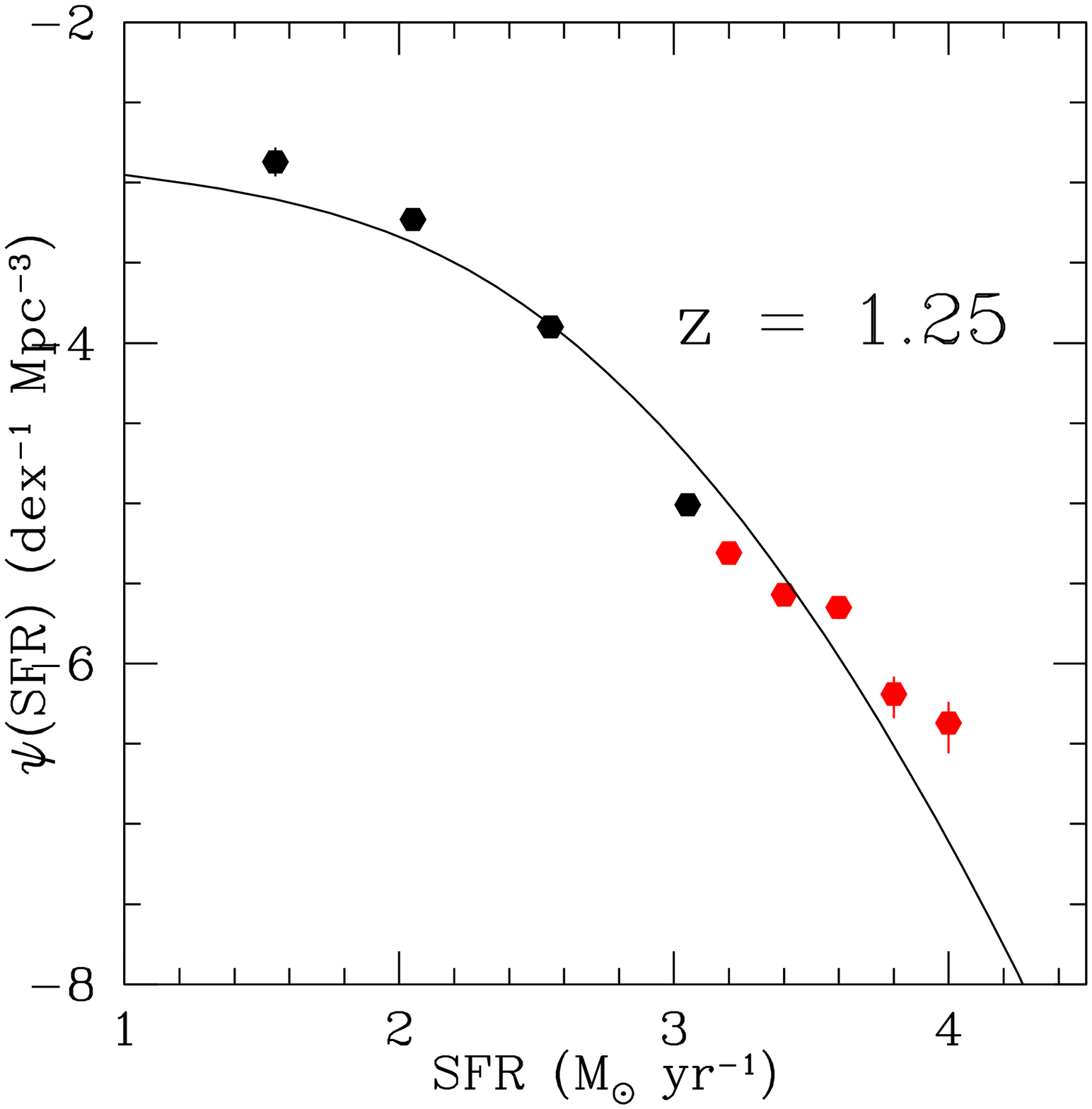}
\includegraphics[width=5.0cm]{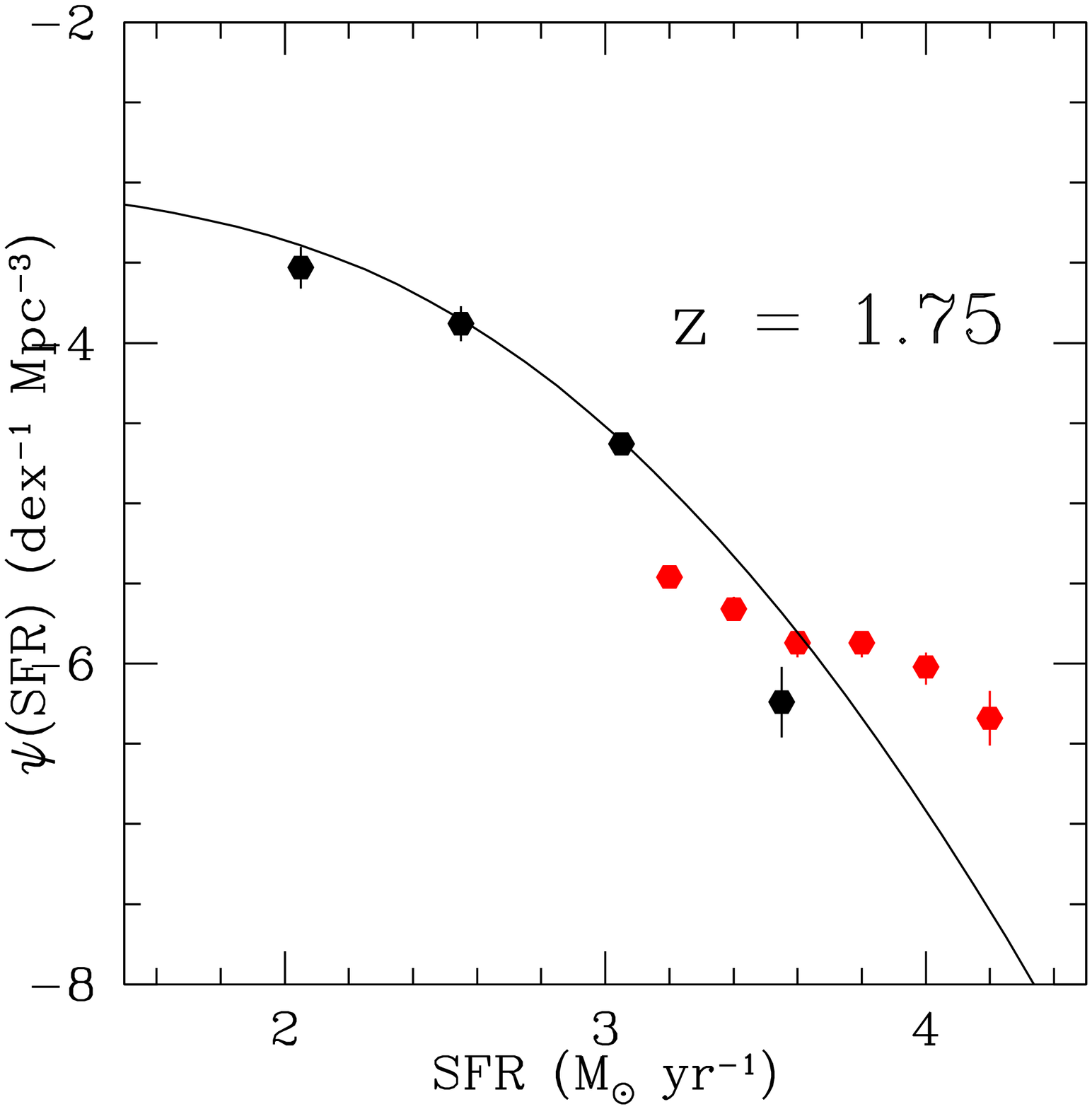}
\includegraphics[width=3.5cm]{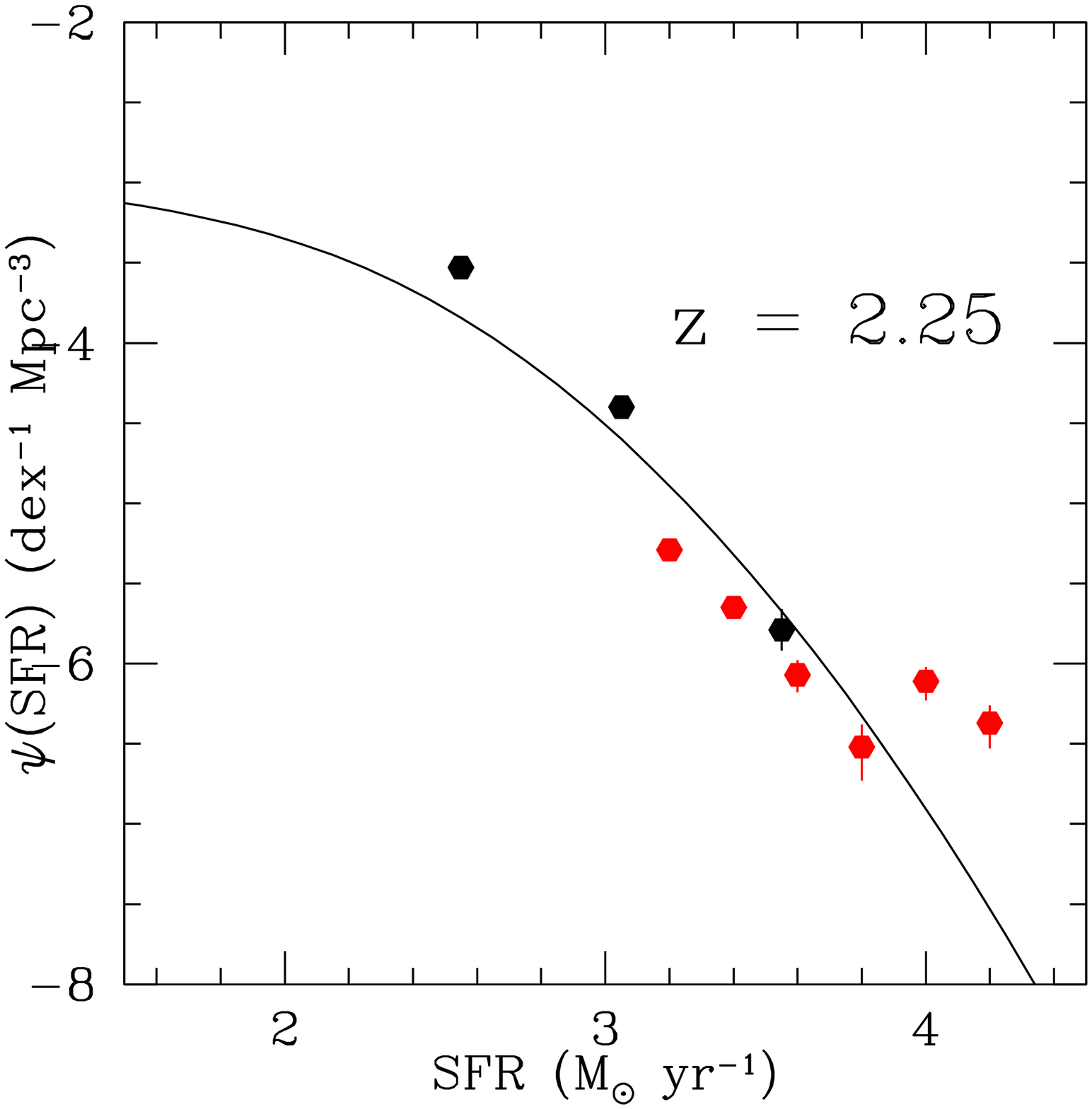}
\includegraphics[width=3.5cm]{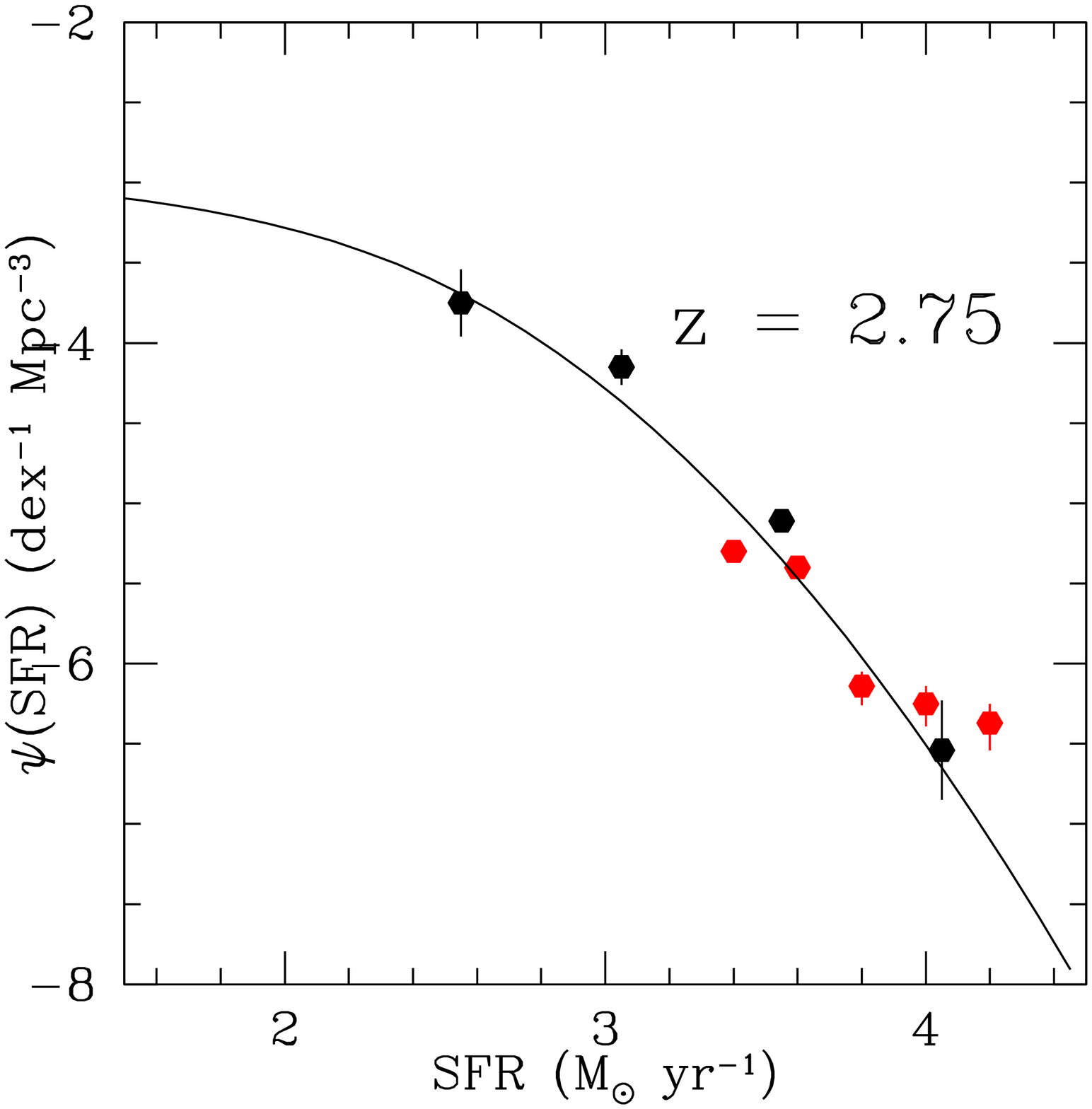}
\includegraphics[width=3.5cm]{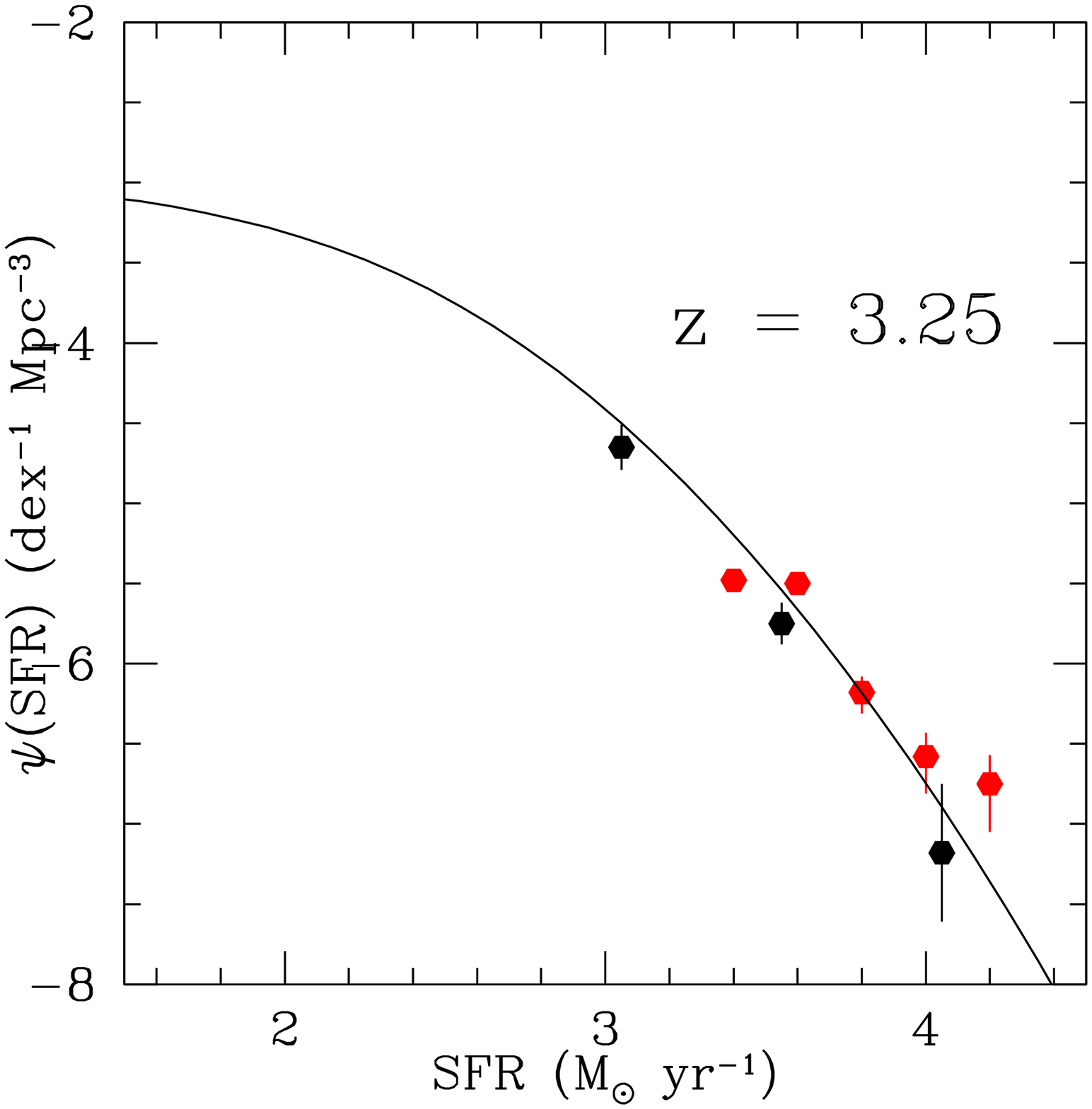}
\includegraphics[width=3.5cm]{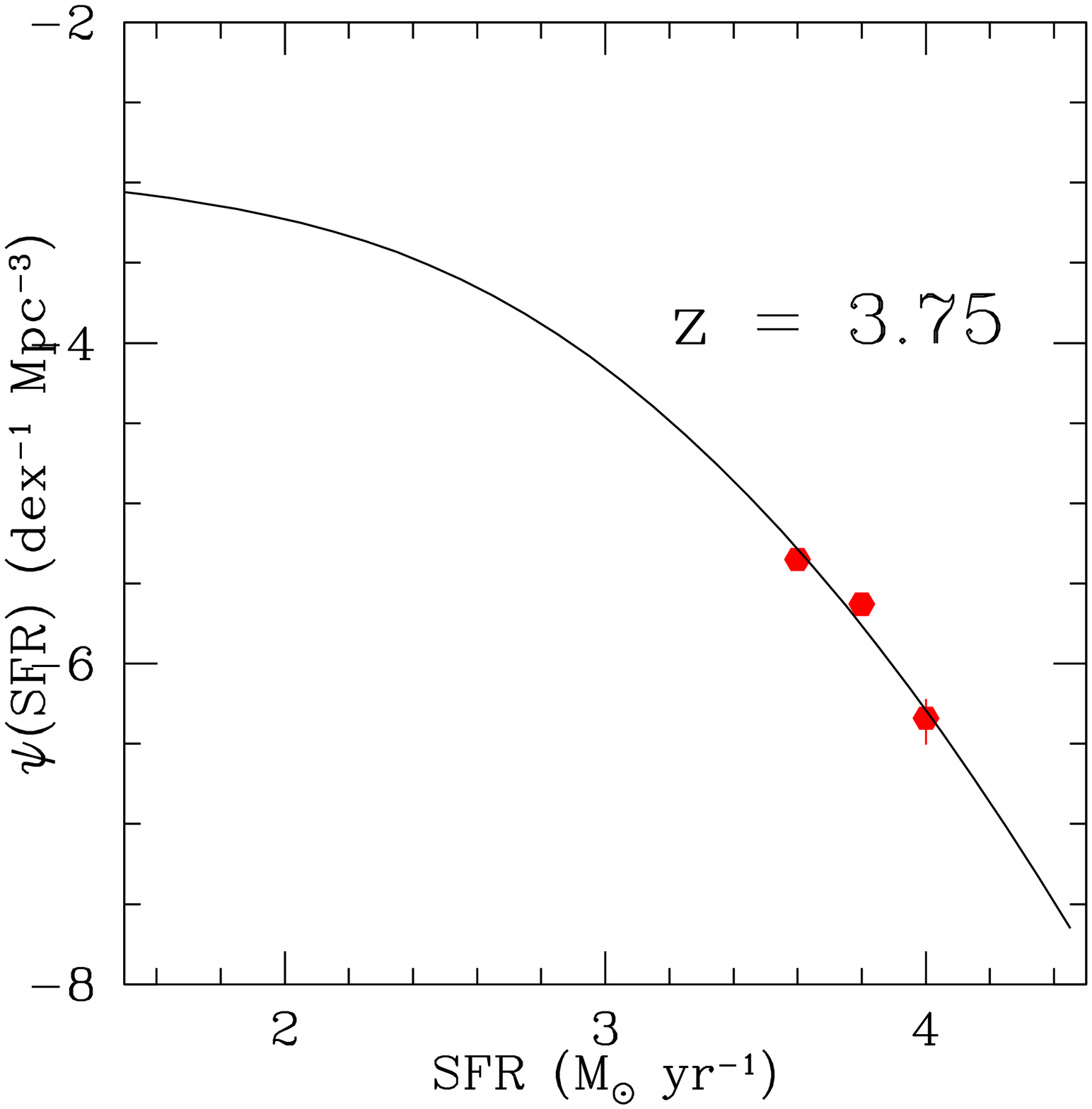}
\includegraphics[width=3.5cm]{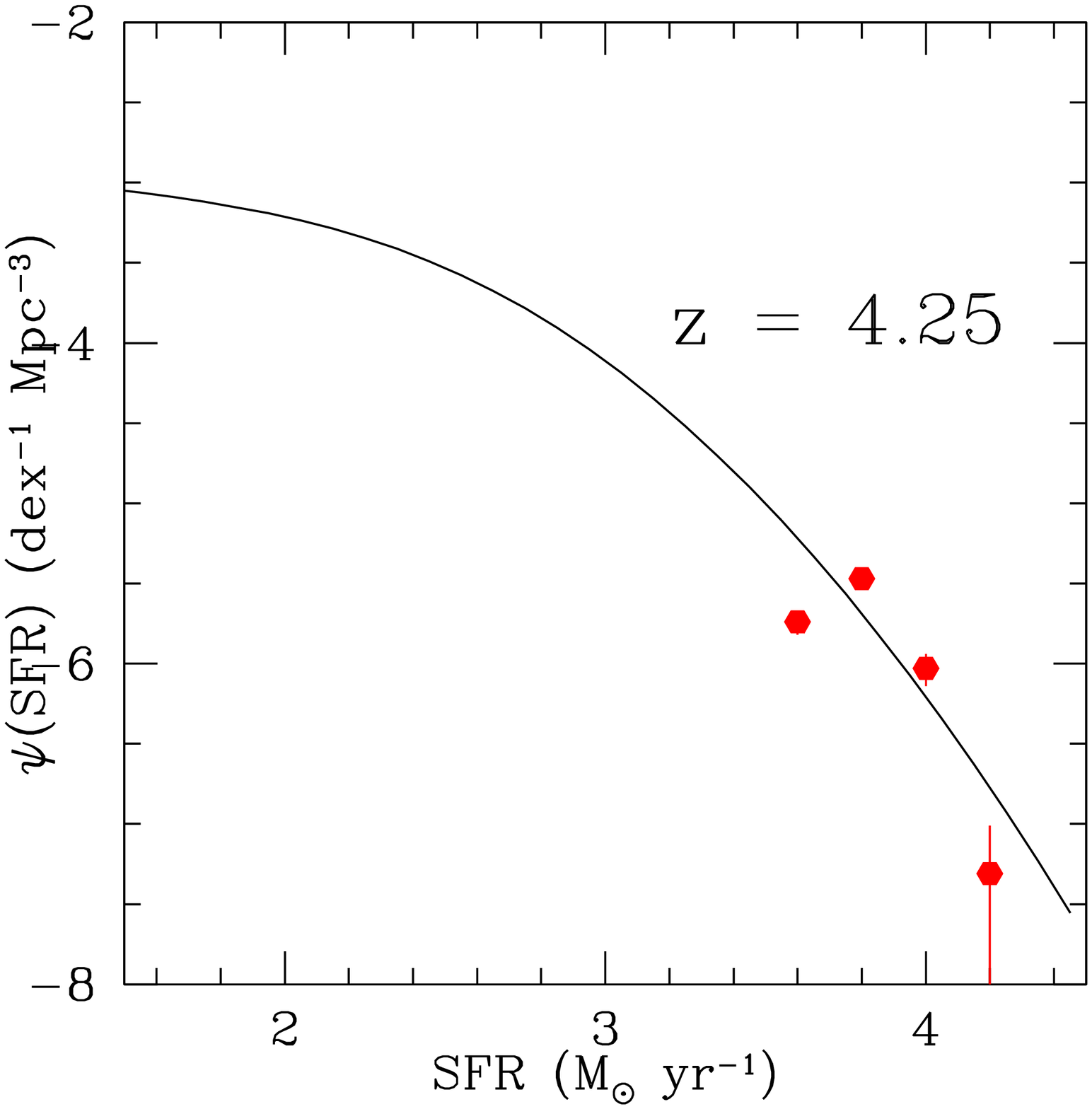}
\includegraphics[width=3.5cm]{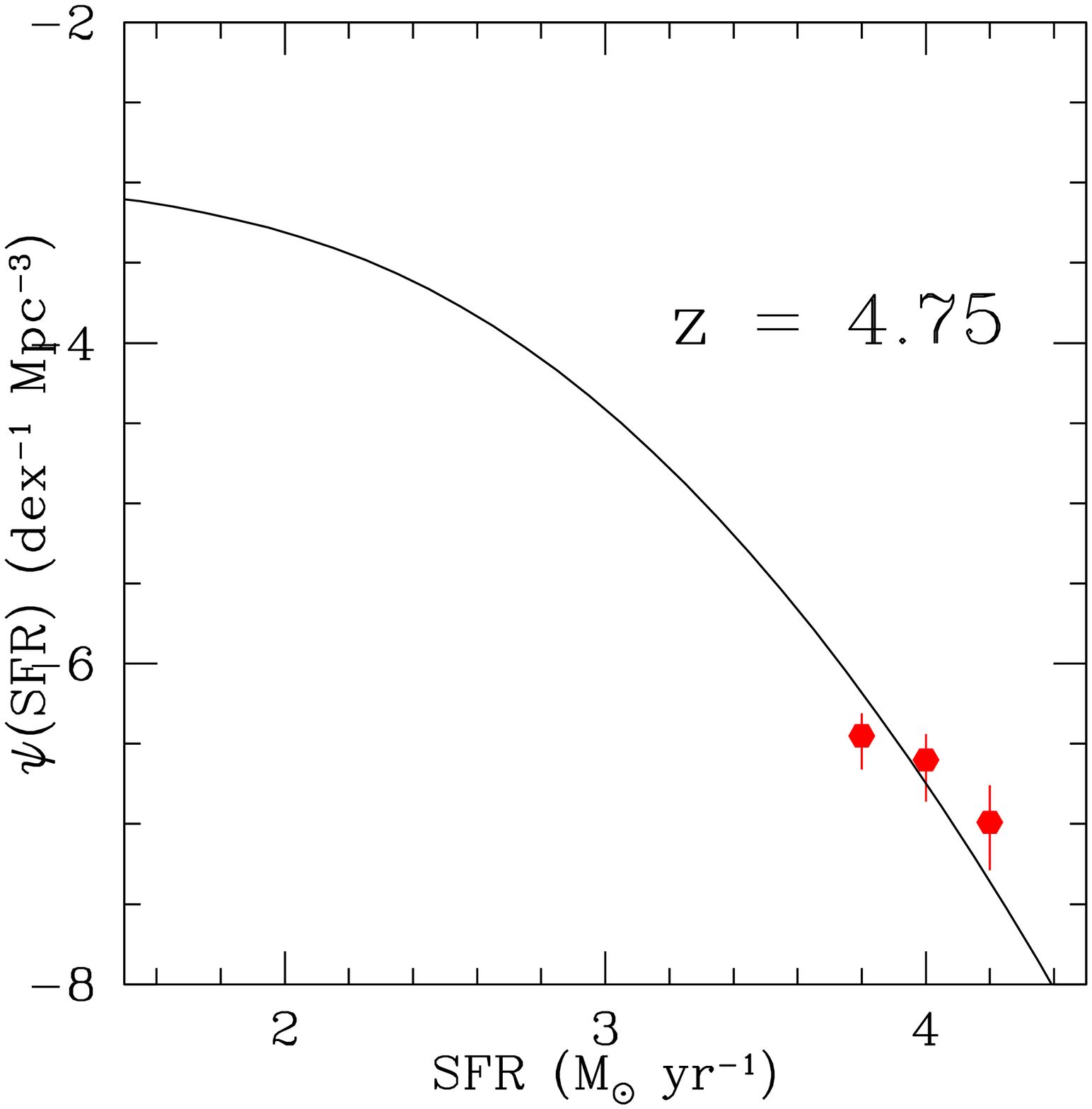}
\includegraphics[width=3.5cm]{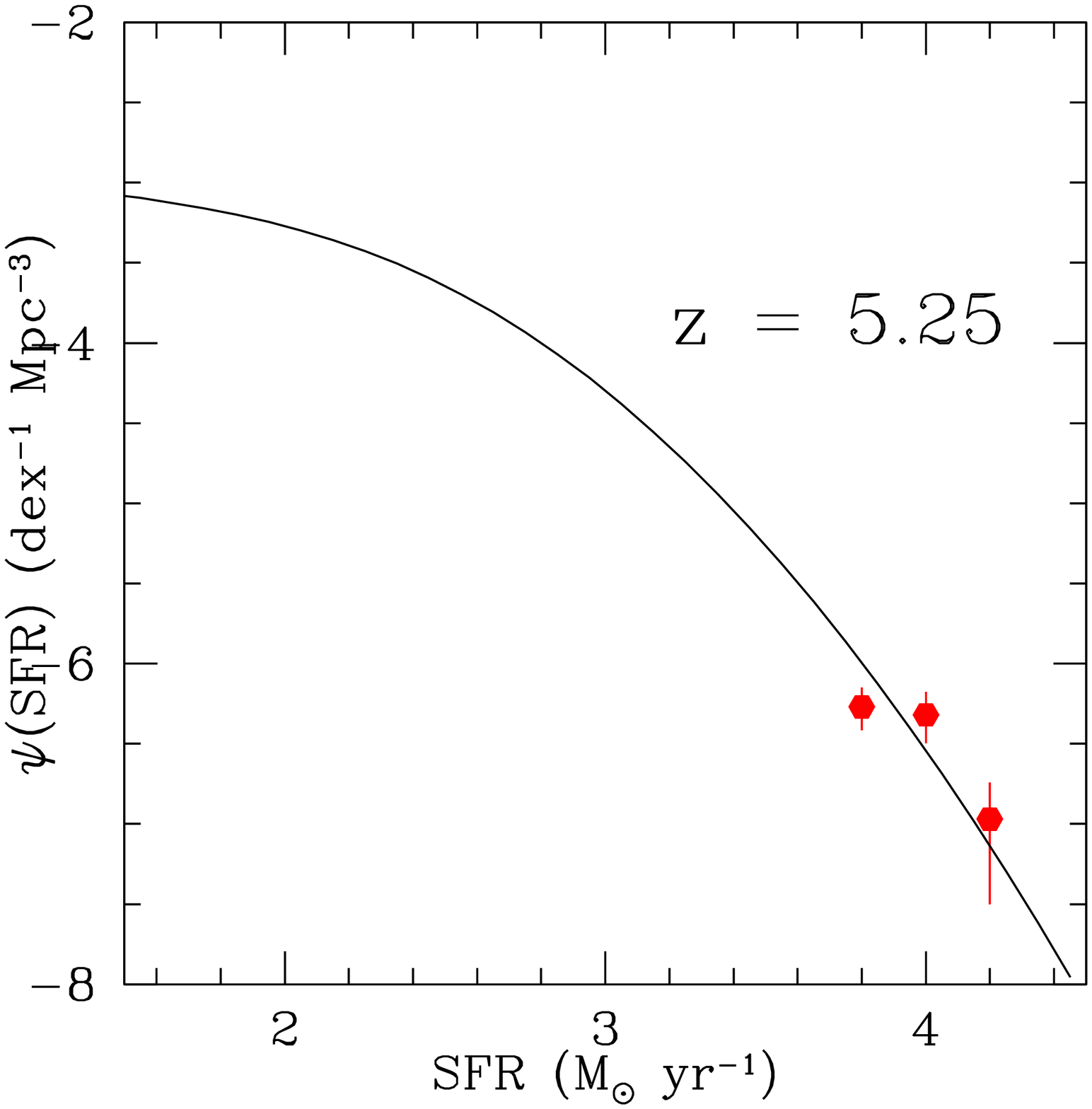}
\includegraphics[width=3.5cm]{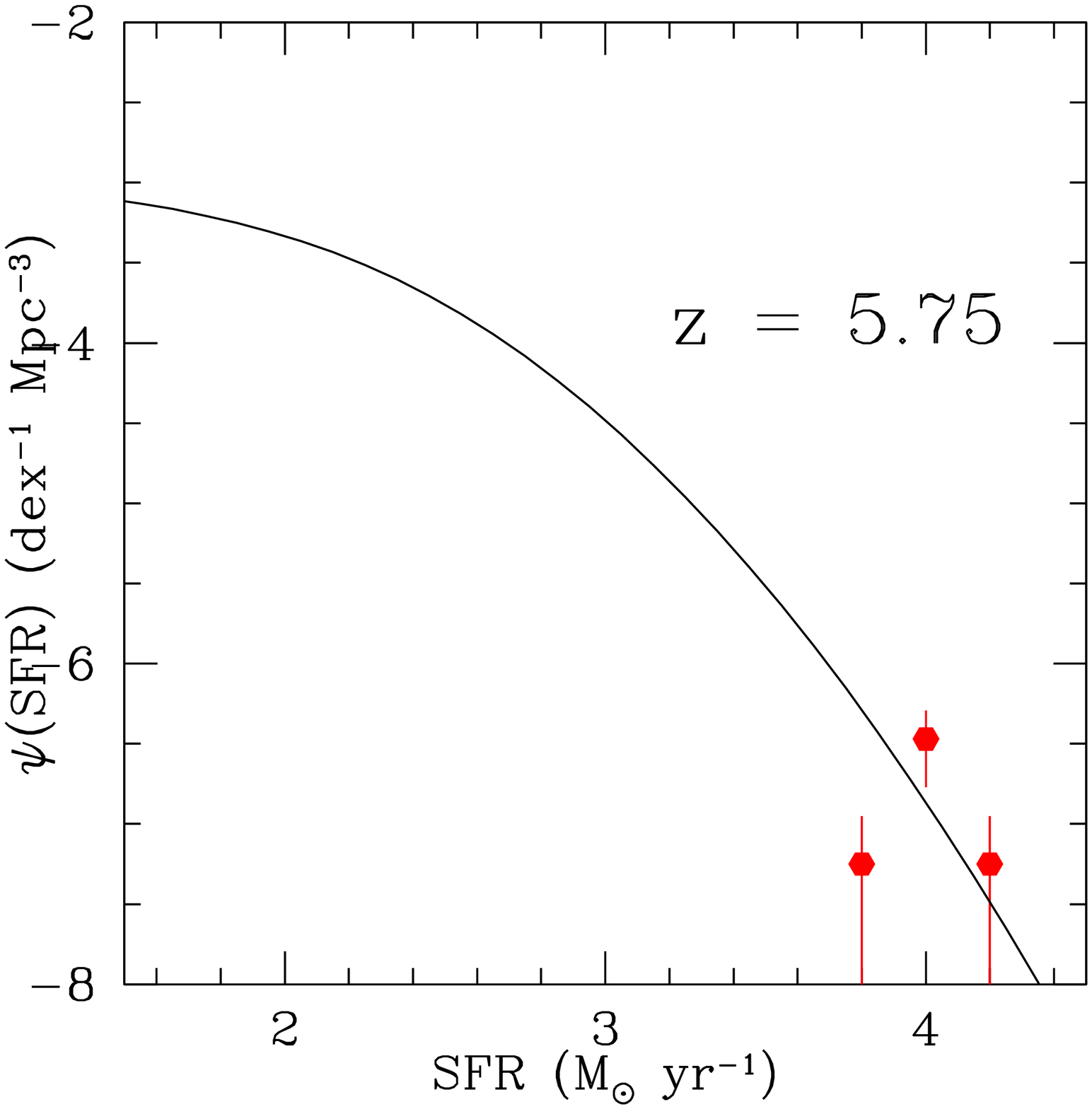}
\caption{
Star formation rate functions for starburst galaxies for z = 0.5-6.0, in bins of 0.5 in redshift. Red points are from present HerMES-SWIRE study, black points are from Gruppioni et al 2013.
}
\end{figure*}

\begin{table*}
\caption{Star-formation rate function parameters}
\begin{tabular}{lllllllllllll}
&&&&&&&&&&&&\\
z = & 0.25 & 0.75 & 1.25 & 1.75 & 2.25 & 2.75 & 3.25 & 3.75 & 4.25 & 4.75 & 5.25 & 5.75\\
&&&&&&&&&&&&\\
{\bf quiescent} &&&&&&&&&&&&\\
&&&&&&&&&&&&\\
no. of gals in fit & 106 & 38 & 41 &&&&&&&&&\\
&&&&&&&&&&&&\\
$log_{10} (\psi_0)$ & -2.43 & [-2.51] & [-2.92] & & & & & & & & & \\
&&&&&&&&&&&&\\
$log_{10} (SFR_0)$ & -0.38 & -0.03 & 0.67 & & & & & & & & &\\
&&&&&&&&&&&&\\
&&&&&&&&&&&&\\
{\bf starburst} &&&&&&&&&&&&\\
&&&&&&&&&&&&\\
no. of gals used in LFs   & 68 & 95 & 295 & 219 & 230 & 252 & 172 & 158 & 89 & 14 & 21 & 5 \\
of which, unassociated & 4 & 6 & 18 & 23 & 130 & 158 & 81 & 132 & 79 & 12 & 20 & 5 \\
&&&&&&&&&&&&\\
$log_{10} (\psi_0)$ & -3.30 & -3.11 & -3.42 & -3.50 & -3.49 & -3.50 & [-3.48] & [-3.48] & [-3.48] & [-3.48] & [-3.48] & [-3.48]\\
&&&&&&&&&&&&\\
$log_{10} (SFR_0)$ & 0.27 & 0.93 & 1.57 & 1.66 & 1.66 & 1.80 & 1.68 & 1.85 & 1.88 & 1.69 & 1.76 & 1.65\\
&&&&&&&&&&&&\\
median no. of & 10 & 8 & 6 & 6 & 6 & 5 & 5 & 5 & 4 & 4 & 4 & 3 \\
photom. bands &&&&&&&&&&&&\\
&&&&&&&&&&&&\\
$log_{10}$ sfrd & -1.90 & -1.28 & -0.95 & -1.06 & -1.05 & -0.82 & -0.99 & -0.82 & -0.79 & -0.99 & -0.92 & -1.03\\
 & $\pm 0.08$ & $\pm 0.21$ & $\pm 0.11$ & $\pm 0.13$ & +0.27 & +0.18 & +0.14 & +0.18 & +0.14 & +0.29 & +0.28 & +0.32\\
 & & & & & -0.09 & -0.09 & -0.41 & -0.36 & -0.41 & -0.46 & -0.44 & -0.38\\
\end{tabular}
\end{table*}

For each bin of redshift and sfr we estimate the contribution to the dust-enshrouded star-formation-rate 
density (sfrd) as n$/$(A V),
where n is the number of sources in the bin, A is the area of the survey (20.3 sq deg), and V is
the co-moving volume sampled by the bin.  For each redshift bin we then fit the star-formation-rate function
with a Saunders functional form (Saunders et al 1990):

$\psi(sfr) = \psi_0.10^{(1-\alpha)(SFR-SFR_0)}$

$\times exp[-(1+10^{(SFR-SFR_0)})^2/(2 \sigma^2)]$
\newline
However because we see only the high luminosity end of the star-formation-rate function we can not  
determine all the parameters of the luminosity function freely from the present samples on their own
in all redshift bins.

For z $<$ 0.5 (Fig 8) we combine our data with the Saunders et al (1990) $\lq$cool$\rq$ and $\lq$warm$\rq$ 60 $\mu$m luminosity functions
as representative of the quiescent and starburst populations, respectively, with a conversion from 60 $\mu$m
luminosity to star-formation rate of sfr = $10^{-9.48} L_{60} M_{\odot} yr^{-1}$, where $L_{60}$ is in solar
units (Rowan-Robinson et al 1997).  
The accuracy of the Saunders et al (1990) luminosity functions has been confirmed in many subsequent
analyses, including the Wang and Rowan-Robinson (2010) determination of the 60 $\mu$m luminosity function
from the IRAS FSS Redshift Catalogue, which gave almost identical results to Saunders et al (1990). Other recent 
determinations of the far infrared and submillimetre local luminosity functions include Vaccari et al (2010), 
Patel et al (2013) and Marchetti et al (2016). There
is reasonable consistency between our 500$\mu$m selected determination of the star-formation rate function for
starburst and quiescent galaxies and
those determined by conversion from the Saunders et al (1990) ‘warm’ and ‘cool’ 60$\mu$m luminosity functions,
although our estimates appear high for the highest luminosity bins. 
For starburst$/$warm galaxies we find $\alpha$ = 1.2 $\pm$ 0.1, and $\sigma$ = 0.60 $\pm$ 0.03.  For
quiescent$/$cool galaxies we find $\alpha$ = 1.2 $\pm$ 0.1, and $\sigma$ = 0.55 $\pm$ 0.05.
For all other redshift bins we then fix $\alpha = 1.2$ but allow a range $\pm 0.1$ in estimating the uncertainty
in the sfrd (which is the integral of the star-formation-rate function).

For z = 0.5-3.5 we combine our data with the Gruppioni et al (2013) total infrared luminosity function,
using a conversion sfr = $10^{-9.70} L_{ir} M_{\odot} yr^{-1}$ (Rowan-Robinson et al 1997), where $L_{ir}$ is in solar
units.  For some of our redshift bins we had to combine two of Gruppioni et al’s
bins to get the equivalent redshift range. We show their z=3-4.2 function compared to our 3-3.5 function, since
the most of their redshifts fall in that bin.  
There is reasonable consistency between these independent estimates
of the star-formation-rate function in the range of star-formation rates in common. 
 
For quiescent galaxies at z = 0.5-1.5 we used $\sigma$ = 0.55 but allowed a 
range of $\pm$ 0.05 in estimating the
uncertainty in the sfrd.  We also fixed $\psi_0$ at the values determined by Gruppioni et al (2013).  

For starburst galaxies in the range z = 0-3 we
find a best fit of $\sigma = 0.60$. We allow a range in $\sigma$ of $\pm 0.03$ in estimating the
uncertainties of other parameters. $\psi_0$ and $SFR_0$ were allowed to be free parameters.  

For z $>$3.5 we have only the estimates of the star-formation-rate function at very high values of the star-formation-rate
from our samples and so fix $\alpha = 1.2$, $\sigma = 0.60$.  
We also note that for z = 0-3 the parameter
$\psi_0$ is consistent with being constant, with an average value $10^{-3.48} dex^{-1} Mpc^{-3}$, so for 
z $>$ 3 we fix $\psi_0$ at this value.  However in estimating the uncertainty in the integrated sfrd in 
each redshift bin, we allow both $\psi_0$ and $SFR_0$ to vary by $\pm 0.5$ dex (this is illustrated in Fig 10).
Because we have data only at high luminosities and star-formation rates at these redshifts, we have to
apply some constraint to the star-formation-rate function parameters, particularly $\psi_0$, to avoid abrupt changes
in this function.
We have also shown results in Fig 11
assuming that $\psi_0$ varies as $(1+z)^{-1}$, as assumed by Gruppioni et al (2013), open circles in Fig 11.
These give slightly lower values of the sfrd. The issue that we are probing only the high-luminosity
end of the star-formation-rate function at z$>$3 is common to other studies, e.g. Cucciati et al (2012),
Gruppioni et al (2013).

Table 4 summarises the parameters for the star-formation-rate function in each redshift bin, with square brackets 
denoting prior fixed values.  
Figures 8, 9 shows a montage of these star-formation-rate functions, with the fitted Saunders
forms. There is a tendency for the very highest sfr bins to be higher than the fitted curves for z $<$ 3.5
but these are generally based on quite small numbers of sources.

Figure 10L shows the evolution of $SFR_0$ and $\psi_0$ with redshift and fits to these with the functional
form introduced by Rowan-Robinson (2009) in modelling infrared and submillimetre source-counts:

$\psi_0(t) = {a_0 + (1-a_0) exp [Q(1-t/t_0)] ((t/t_0)^P-(t/t(z_f))^P}$

\begin{figure*}
\includegraphics[width=7cm]{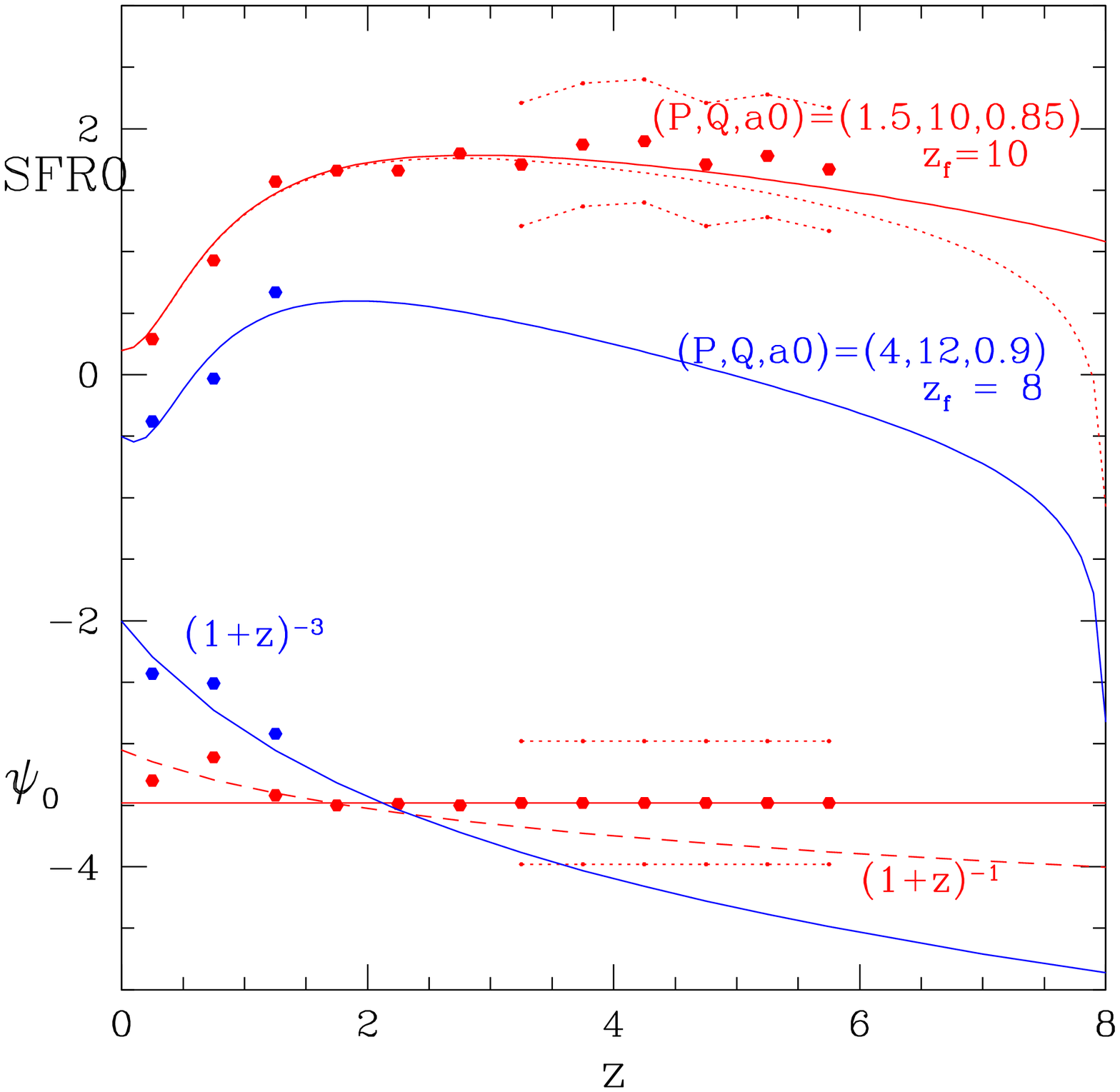}
\includegraphics[width=7cm]{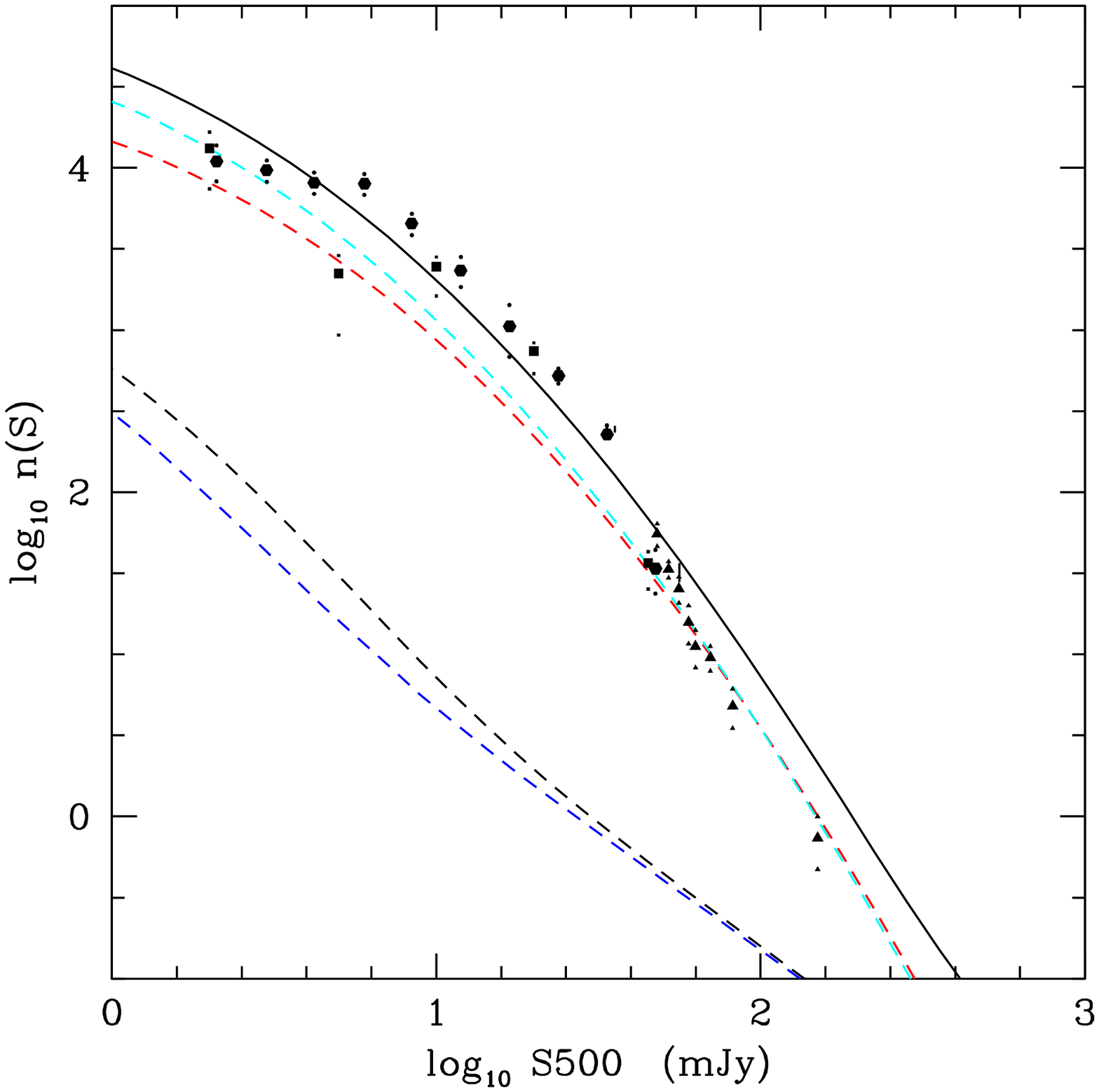}
\caption{L: $\psi_0$ (lower points) and SFR0 (upper points) as a function of redshift for quiescent galaxies (normal cirrus plus cool dust; blue)
and starburst galaxies (M82 + A220 + young starburst; red) as a function of redshift, with simple
analytical fits.
R: Differential 500 $\mu$m counts from HerMES and H-ATLAS surveys (Oliver et al 2010, Glenn et al 2010, 
Clements et al 2010, Bethermin et al 2012) compared with prediction of
evolutionary model derived here. Black dashed locus: cirrus, blue: cool cirrus, red: M82 starburst,
green: A220, black continuous locus: total. 
}
\end{figure*}

As a check we show in Fig 10R a fit to the 500 $\mu$m differential source-counts with these assumed
evolution rates for the different components.  The predicted background intensity at 500 $\mu$m is 
3.3 nW $m^{-2} sr^{-1}$, consistent with direct measurements (2.40$\pm$0.60, Fixsen et al 1998, 
2.70$\pm$0.67 Lagache et al 2000) and stacking analysis (2.80+0.93-0.81 Bethermin et al 2012).

\section{Evolution of the star-formation rate density}

We are now in a position to show the evolution of the star-formation-rate density with redshift
for quiescent galaxies, starbursts, and for the combined total (Fig 11).  This can be compared with the
data compiled by Madau and Dickinson (2014).  The uncertainties we have shown fully take into account
the effect of varying the parameters of the star-formation-rate function, except that at z $>$ 3 we
have not allowed variation of the parameters $\alpha$ and $\sigma$.  To indicate the uncertainties 
associated with the photometric redshifts we have shown a range of redshift corresponding to
the rms uncertainty for the median number of photometric bands in each redshift bin (last line
of table 4), taken from Rowan-Robinson et al (2013, Fig 15).  
Catastrophic photometric redshift
outliers are a negligible problem for z $<$ 1 ($<1.5$\%), and not a big problem for z = 1-2 (few $\%$).
Our analysis combining 0.36-4.5$\mu$m photometric redshift and 250-350 $\mu$m submillimetre redshift 
estimates for z $>$ 2 galaxies identified 27 catastrophic outliers out of 388 objects (7$\%$), and we 
substituted $z_{comb}$ for $z_{phot}$ for these outliers.  

For the 500 $\mu$m sources unassociated with SWIRE objects, the redshifts, determined only from
250-500 $\mu$m data, are highly uncertain, with an estimated range of values $\pm 30\%$ of (1+z) due to the
range of template types seen in the submillimetre ( and a direct estimate from known spectroscopic
redshifts of $\pm 21\%$).
 Extensive discussions of finding high redshift (z$>$4)
galaxies purely from submillimetre photometry have been given by Dowell et al (2014) and Asboth et al (2015).
 The situation can be improved in the future by submillimetre
spectroscopy of these sources.  We believe the redshift uncertainties are unlikely to be responsible
for the high star-formation-rate density we find at z = 3-6. 

\begin{figure*}
\includegraphics[width=14cm]{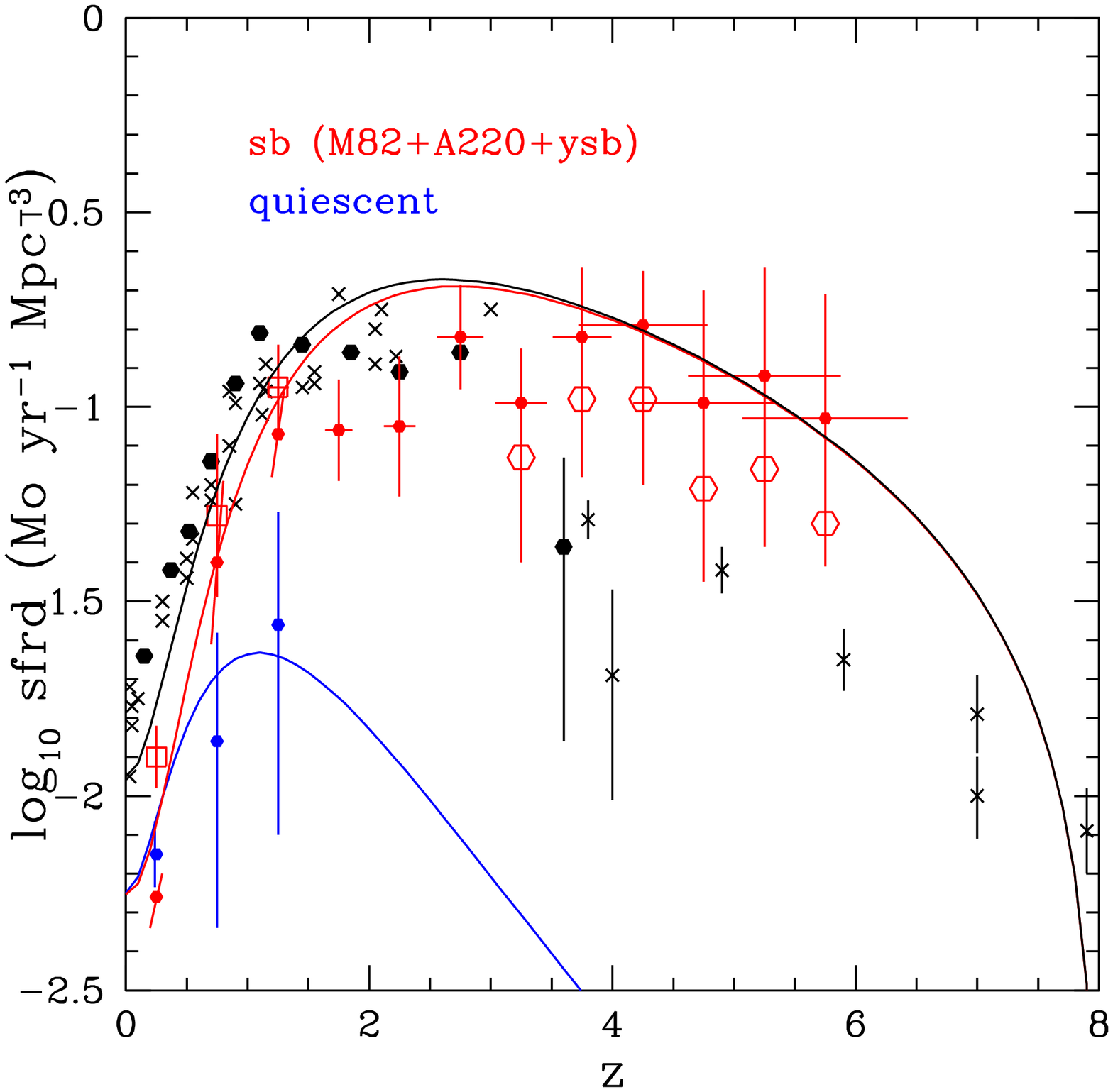}
\caption{
Star formation density as a function of redshift.  Crosses: optical and ultraviolet data summarised
by Madau and Dickinson (2014); black filled hexagons: far infrared data of Gruppioni et al (2013);
filled red hexagons: starburst galaxies from present work derived from 500 $\mu$m sample; filled 
blue hexagons: quiescent galaxies from present work; open squares: sum of starburst and quiescent
contributions; open circles at $z>3$: effect of assuming $\psi_0$ varies as $(1+z)^{-1}$.
}
\end{figure*}

There is reasonable agreement between our results and those of 
Madau and Dickinson (2014), for z $<$ 3.5. 
Working at 500 $\mu$m allows us to make a good determination of the sfrd at z = 3.5-4.5, and a more
uncertain determination, based on small numbers of galaxies, at z = 4.5-6.  Our star-formation-rate 
density is higher than that of Gruppioni et al (2013) at
z=3-4.5, but the latter is based on very few galaxies and is quoted by the authors as essentially
a lower limit.  Our values are also higher than uv estimates at z = 3.5-6, by a factor 2-3, with the
lower value applying if we assume negative density evolution, with $\psi_0$ varying as $(1+z)^{-1}$.  
Although the uncertainties in the submillimetre estimates mean that we can not rule out the possibility
that the ultraviolet estimates are correct, the higher values of the sfrd that we find could also be 
attributed to the fact that uv estimates are not able to assess the contribution of embedded star-formation.
Note that the uv estimates of Bouwens et al (2012a,b) and Schenker et al (2013) do not include any correction for
extinction by dust, though those of Cucciati et al (2012) do.

\section{Discussion and conclusions}

Our conclusion is that the epoch of high star-formation-rate density, and hence of rapid heavy element formation,
stretches from redshift 4 to redshift 1.  This is a significantly earlier start to the epoch of high
star-formation-rate density than assumed in previous studies. This may pose problems for semi-analytic
models for galaxy formation, which tend to set the epoch of intense star-formation at z = 2 to 1.
Gruppioni et al (2015) have discussed the consistency of the submillimetre luminosity function evolution with
semi-analytic models. Their conclusion that the semi-analytic models underpredict the high star-formation rate
seen in starburst galaxies at z $>$ 2 is strengthened by the results of the present paper.

McDowell et al (2014) have constructed a sample of z $>$ 4 submillimetre galaxies by selecting HerMES sources with 
S(500)$>$S(350)$>$S250).  While only 3 of our Table 3 z $>$ 4 sample satisfy S(500)$>$S(350), it is interesting 
that assuming the submillimetre luminosity function does not evolve from z = 2 to 5 they conclude that that the 
sfrd from z=4-6 is between 
1 and 3 times the uv estimates, depending on which luminosity function they use.  

Luminous star-forming galaxies at redshift 5-6 had significant dust opacities in their star-forming
regions.  If this dust originates in Population 2 stars, this in turn has implications for the epoch
when star-formation commenced, as this must be $\sim$ 1 Gyr before z = 6.  Michalowski et al (2010)
have emphasised the problems posed by some high-redshift submillimetre galaxies for the
assumption of AGB dust formation.

Our estimates of the star-formation rate density at z $>$ 4 are based entirely on very exceptional
objects, with star-formation rates $>$ 3,000 $M_{\odot} yr^{-1}$.  Presumably
the lifetime of such exceptional starbursts is significantly shorter than the $\sim 10^8$ yrs seen
in local starbursts, but they still need to be understood in the context of semi-analytic galaxy formation
models.  We believe we have efficiently
removed lensing objects from our study and that these represent real star-formation rates.  However it
is possible that some of the 500 $\mu$m sources not associated with SWIRE galaxies may be lensed and
this could reduce our sfrd estimates at high redshift. If the fraction of lensed objects amongst the unassociated
 sources was the same as for the Herschel sources associated with SWIRE galaxies, i.e. 11$\%$, then the
star-formation rate density at z$>$4 would need to be reduced by this factor.  This would not alter our conclusions.  

While our estimate of the star-formation rate density is highly uncertain, especially at z $>$ 4.5,
it demonstrates the potential of submillimetre selected samples of galaxies for probing the
high redshift universe and the need for further work, especially spectroscopy and submillimetre imaging,
on these high redshift submillimetre galaxies.

\section{Acknowledgements}

{\it Herschel} is an ESA space observatory with science instruments provided by European-led
Principal Investigator consortia and with important participation from NASA. SPIRE has
been developed by a consortium of institutes led by Cardiff University (UK) and
including Univ. Lethbridge (Canada); NAOC (China); CEA, LAM (France); IFSI, Univ.
Padua (Italy); IAC (Spain); Stockholm Observatory (Sweden); Imperial College London,
RAL, UCL-MSSL, UKATC, Univ. Sussex (UK); and Caltech, JPL, NHSC, Univ. Colorado (USA).
This development has been supported by national funding agencies: CSA (Canada); NAOC
(China); CEA, CNES, CNRS (France); ASI (Italy); MCINN (Spain); SNSB (Sweden); STFC,
UKSA (UK); and NASA (USA).

We thank an anonymous referee for helpful comments.


\end{document}